%% file: Bdwarfs.tex
\documentclass[longauth]{aa}  

\usepackage{graphicx}
\usepackage{txfonts}
\usepackage[colorlinks=true,
            citecolor=TealBlue,
            linkcolor=Peach,
            urlcolor =blue]{hyperref}
\usepackage{xspace}
\usepackage[dvipsnames]{xcolor}
\usepackage{amsmath}	
\usepackage{booktabs}
\usepackage{xparse}
\usepackage{caption}
\usepackage{placeins}
\usepackage{cuted}

\newcommand{\bloem}{BLOeM\xspace}
\newcommand{\kms}{\,km\,s$^{-1}$\xspace} 
\newcommand{\Msun}{$\,{\rm M}_\odot$\xspace}  
\newcommand{\Zsun}{$\,{\rm Z}_\odot$\xspace}  
\newcommand{\Teff}{$T_{\rm eff}$\xspace} %
\newcommand{\logg}{$\log g$\xspace} %
\newcommand{\drv}{$\Delta$RV\xspace} %
\newcommand{\drvmax}{$\Delta{\rm RV}_{\rm max}$\xspace}
\newcommand{\rvvar}{RV\,var\xspace}
\newcommand{\pdet}{$p_{\rm det}$\xspace}
\newcommand\spline[3]{\text{#1\,\textsc{\lowercase{#2}}\,$\lambda$#3}}

\NewDocumentCommand{\fbin}{mmg}
{%
    \IfValueTF{#3}
        {%
            $f_{\rm mult} = #1^{+#2}_{-#3}$\% %
        }
        {%
            $f_{\rm mult} = #1 \pm #2$\% %
        }%
}
\NewDocumentCommand{\fobs}{mmg}
{%
    \IfValueTF{#3}
        {%
            $f^{\rm obs}_{\rm mult} = #1^{+#2}_{-#3}$\% %
        }
        {%
            $f^{\rm obs}_{\rm mult} = #1 \pm #2$\% %
        }%
}

\makeatletter
\renewcommand*\aa@pageof{, page \thepage{} of \pageref*{LastPage}} 
\makeatother

\begin{document}

   \title{Binarity at LOw Metallicity (BLOeM):\thanks{Based on observations collected at the European Southern Observatory under ESO programme ID 112.25W2.}}

   \subtitle{Enhanced multiplicity of early B-type dwarfs and giants at $Z=0.2$\Zsun}

   \author{J.~I.\ Villaseñor \inst{\ref{inst:mpia}}
        \and H.\ Sana\inst{\ref{inst:kul}}
        \and L.\ Mahy \inst{\ref{inst:rob}}
        \and T.\ Shenar\inst{\ref{inst:TelAv}}
        \and J.\ Bodensteiner\inst{\ref{inst:eso}}
        \and N.\ Britavskiy \inst{\ref{inst:rob}}          
        \and D.\ J.\ Lennon\inst{\ref{inst:iac}}
        \and M.\ Moe\inst{\ref{inst:wyoming}}
        \and L.\ R.\ Patrick\inst{\ref{inst:cab}}
        \and M.\ Pawlak\inst{\ref{inst:lund}}
        \and D.\ M.\ Bowman\inst{\ref{inst:newcastle}, \ref{inst:kul}}
        \and P.\ A.\ Crowther\inst{\ref{inst:sheffield}}
        \and S.\ E.\ de Mink\inst{\ref{inst:mpa}}
        \and K. Deshmukh \inst{\ref{inst:kul}}
        \and C.\ J.\ Evans \inst{\ref{inst:esa_stsci}}
        \and M.~Fabry \inst{\ref{inst:villanova}, \ref{inst:kul}}
        \and M.\ Fouesneau\inst{\ref{inst:mpia}}
        \and A.\ Herrero \inst{\ref{inst:iac}}
        \and G.\ Holgado\inst{\ref{inst:iac}}
        \and N.\ Langer\inst{\ref{inst:bonn}}
        \and J. Ma\'{\i}z Apell\'aniz\inst{\ref{inst:CAB}}
        \and I.\ Mandel\inst{\ref{inst:monash},\ref{inst:ozgrav}}
        \and L.\ M.\ Oskinova \inst{\ref{inst:up}}
        \and D.\ Pauli\inst{\ref{inst:kul}}
        \and V.~Ramachandran\inst{\ref{inst:ari}}
        \and M.\ Renzo\inst{\ref{inst:AZ}}
        \and H.-W.\ Rix\inst{\ref{inst:mpia}}
        \and D.~F.~Rocha\inst{\ref{inst:ON_Br}}
        \and A.\ A.\ C.\ Sander\inst{\ref{inst:ari}}
        \and F.\ R.\ N.\ Schneider\inst{\ref{inst:hits},\ref{inst:ari}}
        \and K.\ Sen\inst{\ref{inst:umk}, \ref{inst:AZ}}
        \and S.\ Sim\'on-D\'iaz\inst{\ref{inst:iac}}
        \and J.\ Th.\ van Loon\inst{\ref{inst:keele}}
        \and S.~Toonen\inst{\ref{inst:antonpannekoek}}
        \and J.\ S.\ Vink\inst{\ref{inst:armagh}}
          }

   \institute{Max-Planck-Institut für Astronomie, Königstuhl 17, D-69117 Heidelberg, Germany\label{inst:mpia}\\ \email{villasenor@mpia.de}
         \and {Institute of Astronomy, KU Leuven, Celestijnenlaan 200D, 3001 Leuven, Belgium\label{inst:kul}}
        \and {Royal Observatory of Belgium, Avenue Circulaire/Ringlaan 3, B-1180 Brussels, Belgium} \label{inst:rob}
        \and {The School of Physics and Astronomy, Tel Aviv University, Tel Aviv 6997801, Israel\label{inst:TelAv}};
        \and {ESO - European Southern Observatory, Karl-Schwarzschild-Strasse 2, 85748 Garching bei M\"unchen, Germany \label{inst:eso}}
        \and {Instituto de Astrof\'{\i}sica de Canarias, C. V\'{\i}a L\'actea, s/n, 38205 La Laguna, Santa Cruz de Tenerife, Spain\label{inst:iac}}
        \and {University of Wyoming, Physics \& Astronomy Department, 1000~E.~University~Ave., Laramie, WY 82071, USA\label{inst:wyoming}}
        \and {Centro de Astrobiolog\'{\i}a (CSIC-INTA), Ctra.\ Torrej\'on a Ajalvir km 4, 28850 Torrej\'on de Ardoz, Spain\label{inst:cab}}
        \and {Lund Observatory, Division of Astrophysics, Department of Physics, Lund University, Box 43, SE-221 00, Lund, Sweden}\label{inst:lund}
        \and {School of Mathematics, Statistics and Physics, Newcastle University, Newcastle upon Tyne, NE1 7RU, UK\label{inst:newcastle}}
        \and {School of Mathematical and Physical Sciences, University of Sheffield, Hicks Building, Hounsfield Road, Sheffield, S3 7RH, UK\label{inst:sheffield}}
        \and {Max-Planck-Institute for Astrophysics, Karl-Schwarzschild-Strasse 1, 85748 Garching, Germany\label{inst:mpa}}
        \and {European Space Agency (ESA), ESA Office, Space Telescope Science Institute, 3700 San Martin Drive, Baltimore, MD 21218, USA\label{inst:esa_stsci}}
        \and {Villanova University, 800 E.~Lancaster Ave., Villanova, PA 19085, USA\label{inst:villanova}}
        \and {Argelander-Institut f\"{u}r Astronomie, Universit\"{a}t Bonn, Auf dem H\"{u}gel 71, 53121 Bonn, Germany\label{inst:bonn}}
        \and {Centro de Astrobiolog\'{\i}a (CSIC-INTA). Campus ESAC, camino bajo del castillo s/n, 28\,692 Villanueva de la Ca\~nada, Spain\label{inst:CAB}}
        \and {{School of Physics and Astronomy, Monash University, Clayton VIC 3800, Australia}\label{inst:monash}}
        \and {{ARC Centre of Excellence for Gravitational-wave Discovery (OzGrav), Melbourne, Australia}\label{inst:ozgrav}}
        \and {Institut f\"ur Physik und Astronomie, Universit\"at Potsdam, Karl-Liebknecht-Str. 24/25, 14476 Potsdam, Germany\label{inst:up}}
        \and Zentrum f{\"u}r Astronomie der Universit{\"a}t Heidelberg, Astronomisches Rechen-Institut, M{\"o}nchhofstr.\ 12-14, 69120 Heidelberg, Germany\label{inst:ari}
        \and {Department of Astronomy \& Steward Observatory, 933 N.\ Cherry Ave., Tucson, AZ 85721, USA\label{inst:AZ}}
        \and {Observat\'orio Nacional, R. Gen. Jos\'e Cristino, 77 - Vasco da Gama, Rio de Janeiro - RJ, 20921-400, Brazil\label{inst:ON_Br}}
        \and Heidelberger Institut f{\"u}r Theoretische Studien, Schloss-Wolfsbrunnenweg 35, 69118 Heidelberg, Germany\label{inst:hits}
        \and {Institute of Astronomy, Faculty of Physics, Astronomy and Informatics, Nicolaus Copernicus University, Grudziadzka 5, 87-100 Torun, Poland\label{inst:umk}}
        \and {Lennard-Jones Laboratories, Keele University, ST5 5BG, UK\label{inst:keele}}
        \and {{Anton Pannekoek Institute for Astronomy, University of Amsterdam, Science Park 904, 1098 XH Amsterdam, the Netherlands} \label{inst:antonpannekoek}}
        \and {Armagh Observatory, College Hill, Armagh, BT61 9DG, Northern Ireland, UK\label{inst:armagh}}
             }

   \date{Received ; accepted }
 
  \abstract
{Early B-type stars with initial masses between 8 and 15\Msun are frequently found in multiple systems, as is evidenced by multi-epoch spectroscopic campaigns in the Milky Way and the Large Magellanic Cloud (LMC). Previous studies have shown no strong metallicity dependence in the close-binary ($a<10$\,au) fraction or orbital-period distributions between the Milky Way's solar metallicity (\Zsun) and that of the LMC ($Z=0.5$\Zsun). However, similar analyses for a large sample of massive stars in more metal-poor environments are still scarce.
We focus on 309 early B-type stars (luminosity classes III–V) from the Binarity at LOw Metallicity (\bloem) campaign, which targeted nearly 1000 massive stars in the Small Magellanic Cloud (SMC, $Z=0.2$\Zsun) using VLT/FLAMES multi-epoch spectroscopy.
By applying binary detection criteria consistent with previous works on Galactic and LMC samples, we identify 153 stars (91 SB1, 59 SB2, 3 SB3) exhibiting significant radial-velocity (RV) variations, resulting in an observed multiplicity fraction of \fobs{50}{3}. Using Monte Carlo simulations to account for observational biases, we infer an intrinsic close-binary fraction of \fbin{80}{8}. This fraction reduces to $\sim55$\% when increasing our RV threshold from 20 to 80\kms; however, an independent Markov chain Monte Carlo analysis of the peak-to-peak RV distribution (\drvmax) confirms a high multiplicity fraction of \fbin{79}{5}. These findings suggest a possible anti-correlation between metallicity and the fraction of close B-type binaries, with the SMC multiplicity fraction significantly exceeding previous measurements in the LMC and the Galaxy.
The enhanced fraction of close binaries at SMC's low metallicity may have broad implications for massive-star evolution in the early Universe. More frequent mass transfer and envelope stripping could boost the production of exotic transients, stripped supernovae, gravitational-wave progenitors, and sustained UV ionising flux, potentially affecting cosmic reionisation. Theoretical predictions of binary evolution under metal-poor conditions will provide a key test of our results.}

   \keywords{binaries: spectroscopic -- binaries: close -- stars: early-type -- stars: massive -- Magellanic Clouds}

   \titlerunning{Enhanced multiplicity of early B-type dwarfs and giants at $Z=0.2$\Zsun}
   \authorrunning{J.~I. Villaseñor et al.}

   \maketitle

\section{Introduction}\label{sec:intro}

Massive stars go through complex evolutionary scenarios that are significantly influenced by their high multiplicity fraction \citep{mason+09, chini+12, sana+14, kobulnicky+14, dunstall+15, moe+distefano17, ramirez-tannus+24} and frequent binary interactions \citep{sana+12, demink+14}. Spectroscopic studies in the Galaxy and the Large Magellanic Cloud (LMC) consistently indicate that more than 50\% of O- and B-type stars are part of binary or multiple systems within the orbital period range where interactions are expected (${\lesssim}3500$\,d). Including longer orbital periods, the overall multiplicity fraction can exceed 90\% for the O-type stars \citep{sota+14, bordier+22}. 
\citet{sana+12} found an intrinsic multiplicity fraction (after correction for observational biases) of \fbin{69}{9} for a sample of 71 O-type stars in young Galactic clusters. In the Cygnus OB2 association, \citet[][and references therein]{kobulnicky+14} presented orbital solutions for 48 massive stars with spectral types ranging from O5 to B2, resulting in an intrinsic multiplicity fraction near 55\% from a sample of 45 O-type and 83 B-type stars. More recently, \citet{banyard+22} studied the B-type content of the young Galactic cluster NGC~6231, determining \fbin{52}{8} from 80 stars covering the entire B-type spectral type range (B0-B9). While further multiplicity studies have been conducted in the Milky Way (MW), many have not been corrected for observational biases \citep[e.g.][]{barba+17, berlanas+20, ritchie+22}, have targeted narrow spectral ranges \citep{abt+90}, focused on peculiar types \citep[e.g. chemically peculiar stars;][]{schoeller+10}, or addressed larger separations \citep{mason+09,sana+14}. These factors hinder direct comparison of multiplicity fractions due to the different observational strategies and biases.

In the 30 Doradus (30 Dor) region of the LMC ($Z_{\rm LMC}=0.5$\Zsun), the VLT-FLAMES Tarantula Survey \citep[VFTS;][]{evans+11} performed multiplicity studies on 360 O-type \citep{sana+13} and 408 B-type stars \citep[][of which 361 were main-sequence (MS) and giant stars]{dunstall+15}, resulting in binary fractions\footnote{Throughout this paper, we use the terms multiplicity fraction and binary fraction interchangeably, and do not include higher-order multiples unless explicitly stated.} of \fbin{51}{4} and \fbin{58}{11}, respectively. The consistency across these studies suggests that the multiplicity properties of massive stars may be independent of metallicity. Furthermore, the distribution of orbital periods is remarkably similar across most of these samples \citep{almeida+17, villasenor+21,banyard+22}, suggesting that O- and B-type stars in multiple systems are formed with similar efficiency and orbital configurations within the metallicity ranges studied so far, emphasising the need for multiplicity studies of massive stars in lower metallicity environments.

In the case of solar-type binaries, a strong anti-correlation between binary fraction and metallicity has been found \citep{badenes+18, moe+19, offner+22}. \citet{moe+19} computed intrinsic binary fractions for five different observed samples comprising spectroscopic and eclipsing binaries (EBs), and found a combined close binary fraction that drops from $53\pm12$\% at [Fe/H]$=-3.0$ to $10\pm3$\% at [Fe/H]$=+0.5$. This stands in stark contrast to the findings for OB-type stars \citep[see also][]{moe+distefano13} and also to solar-type stars in wider orbits \citep[$a\gtrsim200$\,au;][]{el-badry+rix19}. Both \citet{moe+19} and \citet{el-badry+rix19} agree that the formation of wide binaries is governed by turbulent core fragmentation which is independent of metallicity, but, in the case of close binaries, the main formation mechanism is disc fragmentation up to separations of $a\sim200$\,au. \citet{moe+19} suggest that the difference between the sharp increase in the binary fraction of solar-type stars towards low metallicities and the close-to-constant fraction observed in massive stars lies in the disc of massive protostars. At solar metallicity, these discs are already highly unstable and have a high probability of fragmentation, leaving little room to further increase multiplicity at reduced metallicities.

The findings in the MW and LMC suggest a universality in the multiplicity properties of massive stars, and motivate further investigation on whether trends persist at lower metallicity. Since the LMC metallicity is only half solar, it is essential to perform multiplicity studies of massive stars at even lower metallicity. In this context, the Small Magellanic Cloud (SMC) is a unique environment due to its proximity \citep[62\,kpc;][]{graczyk+20}, low metallicity ($Z_{\rm SMC}=0.2$\Zsun) representative of the high-redshift universe \citep{nakajima+23}, and high content of massive stars \citep{evans+04}. Furthermore, theoretical predictions indicate that the winds of massive stars at SMC metallicity are strongly reduced \citep{vink+01,borklund+21}, with empirical evidence showing that they might be even weaker than predicted \citep{ramachandran+19, rickard+22}. If there is a connection between metallicity and the multiplicity properties of massive stars, the SMC offers a prime environment for testing the universality of these properties. For other low-metallicity galaxies in the Local Group, resolving a significant sample of massive stars with the current generation of telescopes is not possible due to their greater distances.

The Binarity at Low Metallicity \citep[\bloem;][]{shenar+24} campaign addresses this knowledge gap by focusing on the multiplicity properties of massive stars ($M_{\rm ini}\gtrsim8$\Msun) within the SMC. \bloem has observed 929 stars in eight different regions of the SMC. The sample has been divided into subsamples based on their spectral types, emission properties, and luminosity classes, enabling an interpretation of the results within the context of each subsample. The five samples are: the O-type stars \citep{sana+24};
the B-type supergiants \citep[spectral types B0-B3,][]{britavskiy+25}; 
the B-type dwarfs and giants (this work); the cooler supergiants \citep[spectral types later than B3,][]{patrick+25}; 
and the emission-line stars \citep[Oe/Be,][]{bodensteiner+25}. 

In this study, we present the first detailed analysis of the multiplicity properties of a large sample of early B-type dwarf and giant stars at the low metallicity of the SMC. In Sect.~\ref{sec:sample}, we introduce the sample of B-type stars and their main properties. Section~\ref{sec:rvs} details our methodology to measure radial velocities (RVs) and our criteria to classify spectroscopic binaries. Our results on the multiplicity properties are presented in Sect.~\ref{sec:mult_fract}, and our search for periodicity is described in Sect.~\ref{sec:Porb}. We discuss our results in a broader context in Sect.~\ref{sec:disc}, and finally provide our conclusions in Sect.~\ref{sec:conclusions}.

\section{The sample}\label{sec:sample}

The B-type dwarfs and giants sample comprises 309 stars with spectral types ranging from B0 to B2.5 and luminosity classes V to III\footnote{We note that O-type primaries with B-type companions are part of the O-type sample \citep{sana+24}.}. Spectral types for the full sample were presented in \citet{shenar+24}. Specifically, our sample consists of all early B-type ``non-supergiant'' objects, including MS (class V), subgiant (class IV), and relatively unevolved giant stars (class III). We expect our sample to primarily consists of MS stars undergoing core-hydrogen burning and stars that have just recently exhausted hydrogen in their cores. Bright giants and supergiants (classes II and I) form a separate early B-type supergiants subsample, which is the focus of \citet{britavskiy+25}. Hereafter, we refer to our sample as the III/V sample. This classification aligns with that used for the 30~Dor B-type stars, where \citet{dunstall+15} defined ``unevolved'' stars as those with surface gravity \logg$>3.3$\,dex and analysed their multiplicity properties separately from the supergiant sample. However, unlike the 30~Dor unevolved sample, our sample is mainly composed of giant stars (see Sect.~\ref{ssec:fbin_var}). In the case of Be stars, we assume that most of them are binary interaction products \citep{bodensteiner+20c, dallas+22, dodd+24}, therefore they will be the subject of a separate work \citep{bodensteiner+25}.

Our limit on the spectral type is given by the \bloem survey magnitude cut-off of $G<16.5$\,mag. A spectral type of around B2--B3 roughly corresponds to the mass limit above which we expect stars to be core-collapse-supernova (CCSN) progenitors ($M\approx8$\Msun) assuming single-star evolution. In this way, we ensure that the sample is representative of the population of CCSN progenitors that might become gravitational-wave sources.

\bloem has currently obtained and reduced nine epochs from the first semester of observations using the FLAMES instrument \citep[Fibre Large Array Multi Element Spectrograph;][]{pasquini+02} in GIRAFFE mode ($R=6200$) on the UT2 unit telescope of the Very Large Telescope (VLT) at the European Southern Observatory (ESO) in Chile. For ten stars in our sample the number of epochs is below nine, fluctuating between three and seven epochs, and in five additional stars one of the epochs has too low a signal-to-noise ratio (${\rm S/N}\lesssim15$), thus only eight epochs were used. Additional epochs are expected by the end of 2024, with a total of 25 epochs planned by the conclusion of the campaign in September 2025. For a complete overview of the sample selection, data reduction, and spectral classification, see \citet{shenar+24}.

\begin{table}
   \caption{Total baseline and minimum cadence of observation for each of the FLAMES fields.}
   \centering
   \label{tab:cadence}
   \begin{tabular}{lrr}
   \toprule
   \toprule
   {} &  Baseline &  Min Cadence \\
   {} &  [d]      &  [d] \\
   \midrule
   Field 1 &           66.10 &            1.16 \\
   Field 2 &           48.02 &            1.04 \\
   Field 3 &           43.03 &            1.02 \\
   Field 4 &           46.95 &            1.92 \\
   Field 5 &           38.97 &            1.08 \\
   Field 6 &           29.93 &            1.04 \\
   Field 7 &           42.04 &            1.09 \\
   Field 8 &           42.05 &            0.02 \\
   \bottomrule
   \end{tabular}
   \end{table}

Each field's central coordinates and the modified Julian dates (MJDs) per observation are provided by \citet[][Table~A.1]{shenar+24}. Here, we present the total baseline of observations and minimum cadence for each field in Table~\ref{tab:cadence}. We have a mean observational baseline of 45\,d and a minimum cadence of 1 to 2\,d. This cadence allows us to constrain most short-period orbits, except for unevolved contact binaries with periods of less than $P_{\rm orb} \sim 1$\,d. However, these are likely to be EBs and therefore detectable by the Optical Gravitational Lensing Experiment \citep[OGLE;][]{pawlak+16} through their light curves (see Sect.~\ref{sec:EB}).

The total baseline allows us to recover periods of up to about 45\,d for most of the fields, and close to 60\,d for Field 1. Given that about 75\% of binaries with early-B primaries and orbital periods below 500\,d are expected to have periods shorter than 40\,d \citep{villasenor+21}, the cadence of the first nine epochs can effectively detect these short-period binaries while also offering a significant total baseline. However, while nine epochs provide sufficient data to robustly estimate the multiplicity fraction and identify interesting systems, they are at the threshold of what is necessary to constrain orbital periods. Accurate periods, orbital solutions, and the distributions of orbital properties for the full sample will be addressed once all 25 epochs are assembled.

Figure~\ref{fig:smc_fields} shows the distribution of the B-type III/V sample in each field observed by \bloem. Field 6 contains the largest number of objects, 71 stars, followed by Field 8 with 53, whereas Field 2 is the least populated with 18 stars. Notably, Field 4 includes NGC~346, the brightest giant H\,{\sc ii} region in the SMC with a rich OB population \citep{massey+89, Evans+06}, however, we have excluded stars in the centre of the region to avoid crowding. Other less studied clusters/H\,{\sc ii} regions are NGC~371/N76 close to the centre of Field 1 \citep{ripepi+14}, NGC~465/N85, NGC~460/N84 and NGC~456/N83 in Field 6 \citep{dapergolas+91, caplan+96} with most stars coming from NGC~465, and NGC~267/N22 is at the centre of Field 3, surrounded by several other H\,{\sc ii} regions \citep[N25, N26, N21, N23;][]{testor+14}, and NGC~261/N12A in the top right \citep{sano+19}.

   \begin{figure}
   \centering
   \includegraphics[width=\hsize]{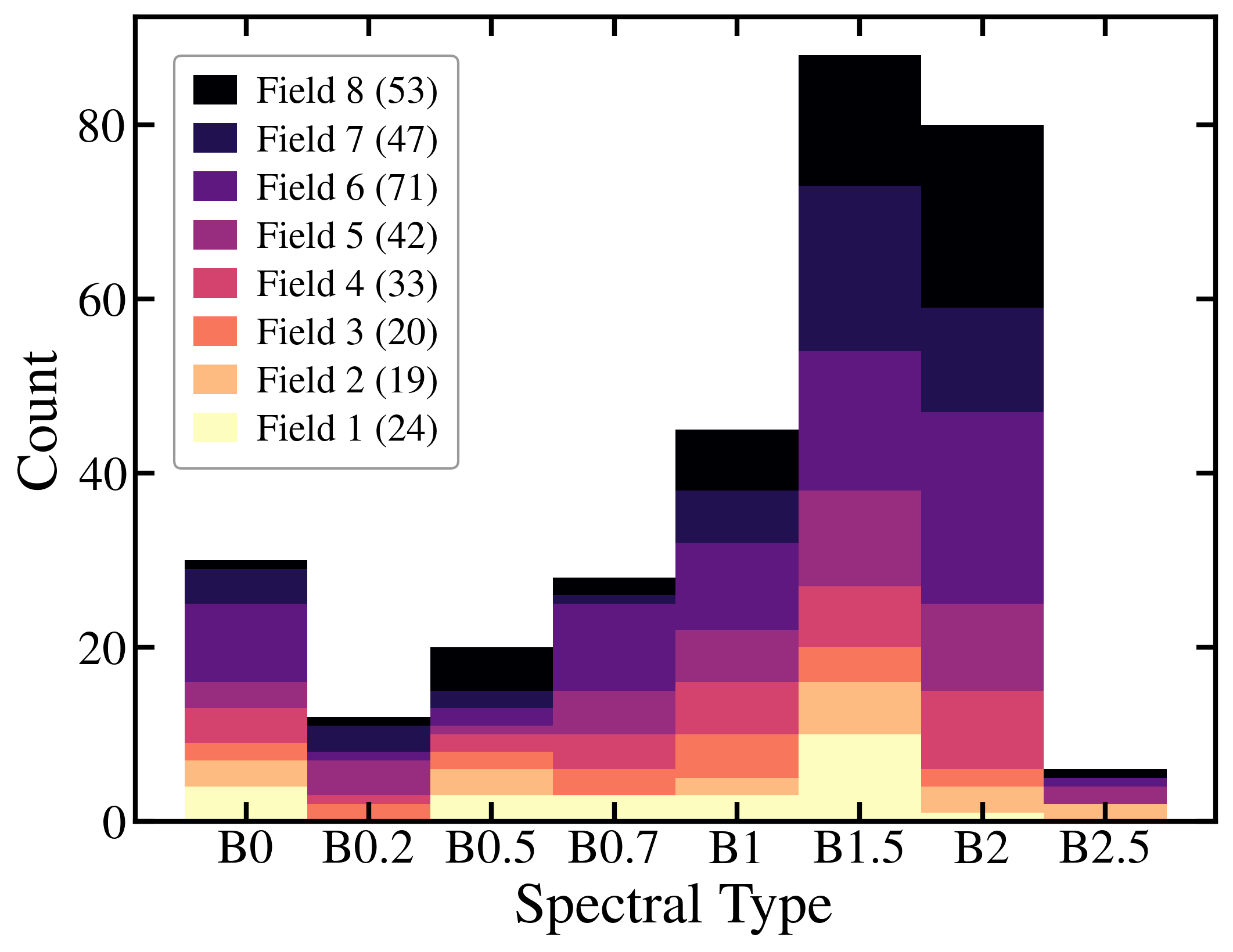}
      \caption{Spectral types of the B-type III/V sample, colour coded by field. The number of stars per field is shown in parentheses.}
      \label{fig:SpTs}
   \end{figure}

The distribution of spectral types per field is shown in Fig.~\ref{fig:SpTs}. The overall distribution is very similar to that found by \citet{ramachandran+19} in the supergiant shell SMC-SGS1 in the Wing of the SMC, a region that was not covered by \bloem. Our sample is dominated by spectral types B1.5 and B2, corresponding roughly to masses between 9 and 12\Msun. There is a good range of spectral types across different fields, but in Fields 7 and 8, the number of B1.5 and B2 seem to be overabundant with respect to the earliest stars (B0--B0.2), potentially suggesting an older population. We computed the ratio of latest-to-earliest type stars for these spectral types (i.e. $N$(B1.5--B2)/$N$(B0--B0.2)), finding ratios of 4.4 and 18.0 for Fields 7 and 8 respectively, and a mean of 2.9 for the other six fields. Performing a Z-test returns that these differences are significant, with a z-score of 5 (Field 7) and 49 (Field 8) and p-values of $5.7\times10^{-7}$ (Field 7) and 0.0 (Field 8). This is consistent with Fields 7 and 8 being areas without prominent star-forming regions (see Fig.~\ref{fig:smc_fields} and details in the main text). In contrast, Field 3, which exhibits several H\,{\sc ii} regions, has the lowest ratio of latest-to-earliest stars with 1.5. We note, however, that an uncertainty of one spectral subtype should be expected. Such uncertainty could at least partially explain the lack of B0.2 and B0.5 stars. These are stars that show weak \ion{He}{ii}, and since \spline{He}{ii}{4686} (stronger than \spline{He}{ii}{4542}) is not covered by the FLAMES LR02 setup, they are more difficult to classify. Given the small number of stars in some of the fields, this variation could also result from sampling effects or constraints in the observational design. 

Examining the star formation history of the SMC \citep[][see their figures 3 and 6]{harris+zaritsky04} we can see that the SMC has had a relatively constant star formation rate (SFR) from 2.5 Gyr until approximately 40 Myr ago, when the SFR started to decrease. All \bloem fields have experienced star-forming episodes in the last 40 to 10 Myr according to the SFR maps of \citet{harris+zaritsky04}. Fields 6, 7, 8 are located in regions that experienced a strong SFR 10 Myr ago, but not in the last 6 to 4 Myr, whereas Field 3 can be associated with strong star formation at 10 and 6 Myr, which is consistent with our suggestion that Field 3 might contain a slightly younger population from the spectral types distribution analysis. Other fields that show a high SFR in the last 4 Myr from the maps of \citet{harris+zaritsky04} are 1 and 4, containing NGC~371 and NGC~346 respectively. We note however that our sample excludes O-type and supergiant stars, so while the presence (or absence) of early B stars can give a qualitative sense of recent star formation, it does not capture the youngest or most evolved populations. Thus, care must be taken when associating these stars alone with specific star-formation epochs. A quantitative analysis of the \bloem stars by Bestenlehner et al. (in prep.) finds median ages for Fields 1--8 in the range of ~8--13 Myr, consistent with our identification of star-formation activity in the last ~10 Myr, but reflecting a wider range of masses and evolutionary stages than those considered here.

\section{Radial velocities}\label{sec:rvs}

\begin{figure}
    \centering
    \includegraphics[width=\linewidth]{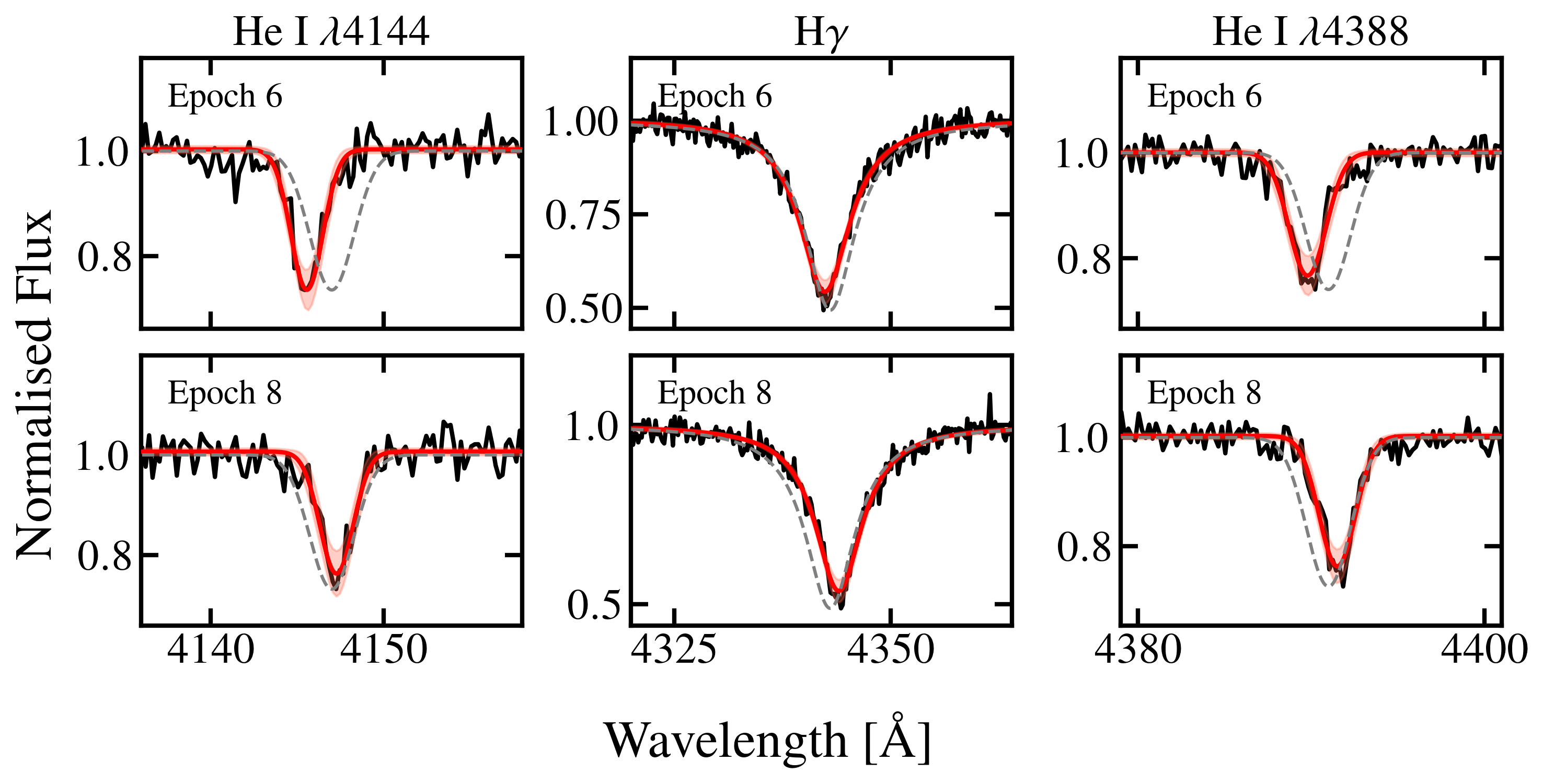}
    \includegraphics[width=\linewidth]{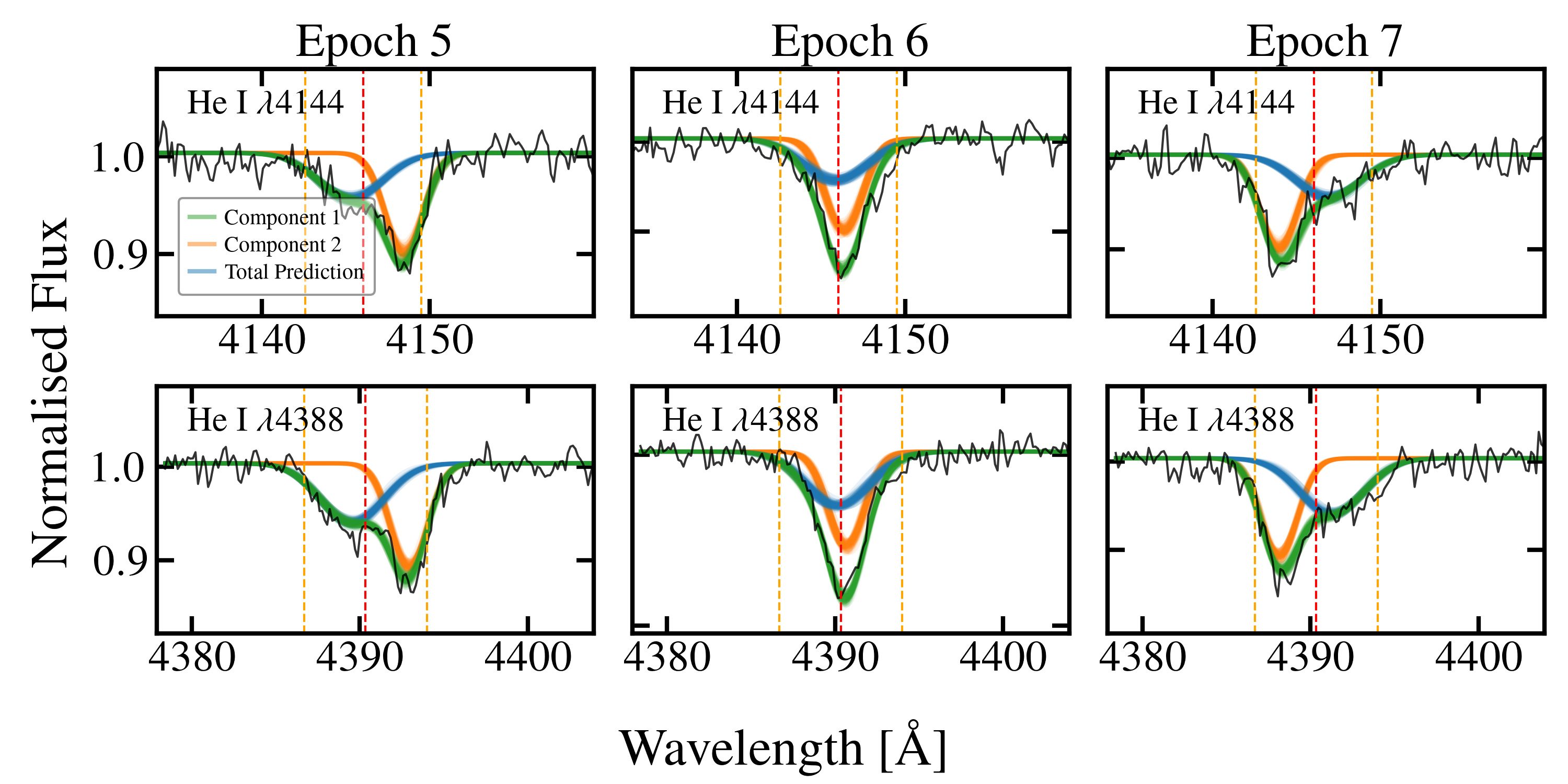}
    \caption{Top: Spectral-line fits for the SB1 system BLOeM~4-097. Three different spectral lines are shown for epochs closest to quadrature. Observed spectrum (black) is overlaid by best fit and $1\sigma$ confidence levels (red); dashed grey profiles indicate initial guesses. Bottom: Similar to above, but for SB2 system BLOeM~1-055. Two \ion{He}{I} lines at three epochs (near quadrature and conjunction) are shown. The dashed red line marks the mean SMC velocity of 172\kms \citep{evans+howarth08}, with a $\pm200$\kms deviation (dashed yellow lines).}
    \label{fig:sb1_fit}
\end{figure}

To measure RVs we have used the same tools as for the B-type stars in 30~Dor \citep{villasenor+21}, which are now part of the MINATO \citep[Massive bINaries Analysis TOols,][]{villasenor+23} {\tt python} repository\footnote{\url{https://github.com/jvillasr/MINATO}}, in a module called \texttt{ravel}. Briefly, it is a spectral line profile fitting tool, which uses Gaussian or Lorentzian (in the case of Balmer lines) profiles to fit a set of spectral lines. In the early B-type stars' spectra, there is a rich set of spectral lines covered by the wavelength range of FLAMES; \spline{He}{I}{4026, 4144, 4388, 4471}, and the Balmer lines H$\delta$ and H$\gamma$. There are also a few metal lines that are usually used in the analysis of later-type stars such as \spline{Si}{ii}{4128-31} and \spline{Mg}{ii}{4481}. However, these are very weak in the stars in our sample due to their spectral range and due to the low metallicity of the SMC, thus we have not included them in our RV measurements. For the earliest types, \spline{He}{ii}{4542} is commonly present, but it is weaker than the \ion{He}{i} lines and therefore provides less accurate RVs. This is similar for the case of \spline{Si}{iii}{4553}, which is usually too shallow for precise RV measurements as a result of the SMC's low metallicity. All mentioned lines are fitted, but \texttt{ravel} selects the ones used to compute a weighted mean RV for each epoch based on the errors of the Gaussian or Lorentzian fit using a median-absolute-deviation \citep[MAD;][]{hampel74} test. For this reason, in most cases, only the \ion{He}{I} and Balmer lines (H$\gamma$, H$\delta$) are used, and \ion{Si}{II} and \ion{Mg}{ii} lines did not pass the MAD test for any of the systems. An example fit for a typical SB1 system is shown in the top plot of Fig.~\ref{fig:sb1_fit}. Two \ion{He}{I} lines and one Balmer line are displayed to illustrate the use of Gaussian and Lorentzian profiles.

\subsection{Double-lined spectroscopic binaries}

A significant improvement in our methodology, compared to the approach used for the 30~Dor population by \citet{villasenor+21}, involves the fitting of double-lined spectroscopic binaries (SB2s). The previous line-by-line fitting method was effective for single-lined binaries (SB1s) but struggled with SB2s, particularly when the binary components had similar flux ratios. This often led to misidentifications and the rejection of affected epochs by the code, severely affecting the number of epochs used in the analysis.

To address these challenges, we have implemented a simultaneous fitting approach using probabilistic programming with \texttt{NumPyro} \citep{bingham2019pyro, phan2019composable}. Similar to the technique used by \citet{sana+13} and \citet{almeida+17}, our method simultaneously fits all selected spectral lines and epochs. The approach allows the RV of each component to vary across epochs; however, all lines from the same component share the RV at a given epoch, while maintaining consistent line profiles (amplitude and width) for each spectral line across epochs. By modelling the data in this simultaneous way, we improve our ability to distinguish and separate the spectral components, even when their flux contributions are comparable.

The probabilistic programming framework offers several advantages \citep[e.g.][]{nightingale+21}: (i) the direct incorporation of uncertainties in the model parameters, which enhances the robustness of our fits without introducing biases;
(ii) comprehensive uncertainty quantification through full posterior distributions for all fitting parameters, enabling a more complete understanding of parameter uncertainties and their correlations;
(iii) efficient data handling by facilitating the simultaneous modelling of multiple spectral lines and epochs, improving computational efficiency and the accuracy of parameter estimations.

These enhancements lead to more reliable identification of SB2 components, improvements in the determination of RV uncertainties, and reduced the need to reject epochs due to misidentifications. However, challenges remain, particularly in cases with limited epochs, contributions from third objects, or significant profile variability. Future work will aim to refine our models to better address these complexities. The bottom plot of Fig.~\ref{fig:sb1_fit} demonstrates the SB2 fitting method by showcasing a double-component fit for the \spline{He}{i}{4144} and \spline{He}{i}{4388} lines of star BLOeM~1-055. The central panel (epoch 6) displays a phase near conjunction, where the two components are fully blended. In contrast, epochs with higher RV variations (e.g. epochs 5 and 7) provide better constraints on the line profiles of the individual components, enabling a reliable fit even for highly blended epochs.

\subsection{Binary criteria}\label{sec:crit}

After weak or absent lines are omitted and epochs with a very low S/N (${\rm S/N}\lesssim15$) are rejected \citep[see details in][]{villasenor+21}, we measure the peak-to-peak RV difference (\drv), which is the maximum difference between any two of the available nine epochs. These measurements are used to classify stars as spectroscopic binaries based on the criteria used by \citet{sana+13, dunstall+15, bodensteiner+21, banyard+22}. The first criterion is based on the peak-to-peak RV variability. Stars with \drv$>C$ are classified as binary candidates, where $C$ is a specified threshold. \citet{dunstall+15} adopted a value of $C=16$\kms for the B-type stars in the LMC, based on their analysis of the binary fraction versus \drv threshold, which showed a distinct break at 16\kms for the B supergiants in their sample. Given that this feature is absent from our results, we adopt a value of $C=20$\kms, consistent with other studies \citep{sana+13, bodensteiner+21, banyard+22}. 

Furthermore, it is well known that B-type stars often exhibit pulsations \citep{Aerts+2009, Bowman2020}. These pulsations can cause intrinsic RV variability with amplitudes peaking in the 10--20\kms range \citep{stankov+handler05, hey+aerts24}. By adopting $C = 20$\kms, we aim to exclude most stars displaying intrinsic variability caused by pulsations, thereby reducing contamination in our binary classification.

The second criterion, introduced by \citet{sana+13}, classifies a system as a spectroscopic binary if, in addition to the first criterion, any pair of its RV measurements satisfies the following condition:

\begin{equation}\label{eq:sigmad}
    \sigma_{ij}=\frac{\left\lvert \varv_i-\varv_j \right\rvert}{\sqrt{\sigma^2_i + \sigma^2_j}} > 4\,{\rm ,}
\end{equation}
where $\varv_i$ and $\varv_j$ are the individual RV measurements taken at two different epochs, and $\sigma_i$ and $\sigma_j$ are the uncertainties associated with each RV measurement. We define $\sigma_{\rm detect}={\rm max}(\sigma_{ij})$, representing the detection significance, which quantifies how significantly different two RV measurements are relative to their combined uncertainties. Since the RV uncertainties are critical for the determination of $\sigma_{\rm detect}$, we test their reliability in Appendix~\ref{append:rverr}. If either of the two criteria are not met, we classify the star as a RV variable (\rvvar) if $\sigma_{\rm detect}>2$, otherwise, we do not consider the variability significant and classify those stars as RV constant (RV\,cst). Stars classified as \rvvar inevitably include unidentified binaries in long-period orbits or with low-mass companions, in addition to stars showing intrinsic RV variability (such as pulsations or atmospheric variability), which may or may not be truly single stars. For this reason, we correct for the observational biases and compute the intrinsic binary fraction (see Sect.~\ref{sec:mult_fract}). 

For the binary classification procedure, we have first identified SB2 systems by visual inspection, and flagged them as SB2s so that \texttt{ravel} computes their RVs adequately. With the RV measurements computed, the binary classification is done automatically according to the two criteria, but some classifications were reevaluated by visual inspection of the fits to spectral line profiles and refitted. In the same way, we have classified three systems as SB3: BLOeM~5-062, BLOeM~6-062, and BLOeM~7-057. We base these classifications on the presence of one or two narrow cores in the \ion{He}{i} lines, with broad profiles on both sides of the narrow lines, or a clear SB2 that cannot be successfully fitted with two Gaussian profiles (see examples in Appendix~\ref{append:sb3}). Furthermore, we found potential signals of contribution from a third star in the spectra of 15 SB2 systems. These have been flagged in the comments of Table~D.1 (see also notes in Appendix~\ref{append:EBs}).

\section{Multiplicity fraction}\label{sec:mult_fract}

\begin{figure}
   \centering
   \includegraphics[width=\hsize]{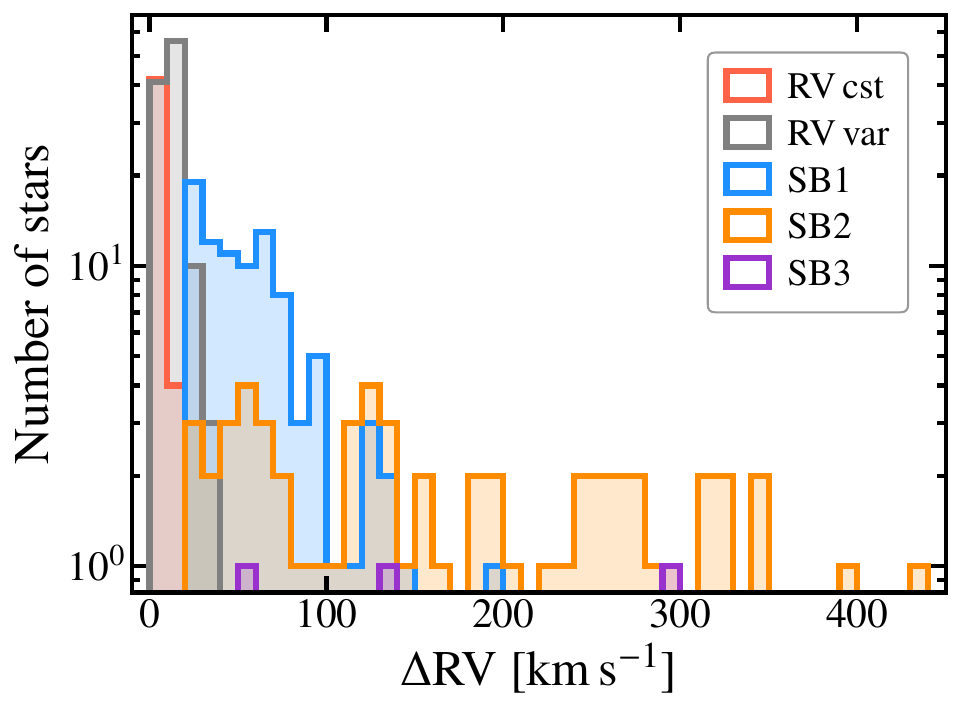}
      \caption{Maximum peak-to-peak RVs for our sample, colour-coded by binary status. We define RV\,cst stars as those with peak-to-peak RV variability not reaching $2\sigma$ significance, which we consider as constant or non-variable. Whereas stars that surpass the $2\sigma$ level but do not meet both multiplicity criteria are considered variable (\rvvar). \drv values are grouped in bins of 10\kms.}
      \label{fig:RVhist}
\end{figure}

Figure~\ref{fig:RVhist} shows the distribution of \drv, for SB1, SB2, SB3, \rvvar, and RV\,cst stars. We classified 46 objects as RV\,cst, of which 42 have \drv below 10\kms. Among a total of 110 \rvvar stars, 13 show \drv higher than 20\kms but do not meet our second binary criterion. All these 13 systems have low S/N, and most of them exhibit line-profile variability that might be attributable to undetected companions or pulsations, which can mimic or blur binary signatures. We also see eight SB2 systems with \drv$<50$\kms, these systems have been clearly identified as SB2 systems through our visual inspection, but their RVs might be affected by the contribution of a third star. In four of these cases, the optically dominant star in the spectrum presents the largest RV variation, suggesting it is the less massive star in the system. 
We define \drv as that of the currently more massive star (that with the smaller RV variations in SB2/3 systems), but in cases where mass transfer has taken place, the dominant star in the spectrum might be the donor of an Algol binary \citep[MS stars with very high luminosity to mass ratio;][]{vanrensbergen+11, sen+22} or a low-mass bloated stripped star \citep[e.g.][]{shenar+20,bodensteiner+20b,villasenor+23, ramachandran+24}, much less massive than the gainer.

\begin{figure}
   \centering
   \includegraphics[width=\linewidth]{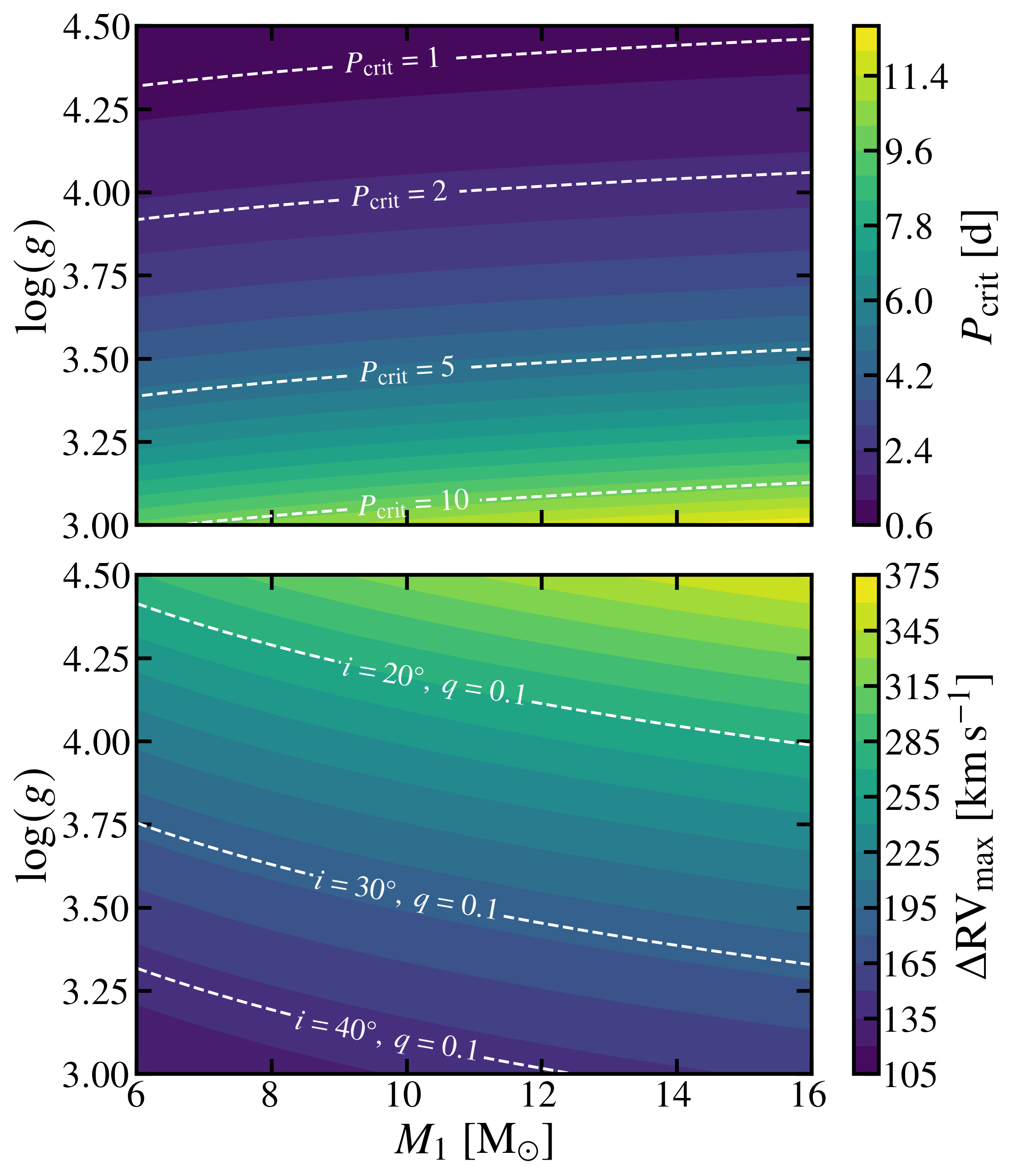}
   \caption{Top panel: Contour plot showing the critical period ($P_{\rm crit}$) as a function of primary mass ($M_1$) and surface gravity ($\log(g)$) for systems with $i=60^\circ$ and $q=0.5$. Dashed white contour lines highlight selected $P_{\rm crit}$ values (1, 2, 5, and 10 days).
   Bottom Panel: As the top panel, this time displaying the corresponding change in RV ($\Delta \mathrm{RV}_{\rm max}$) computed at $P_{\rm crit}$. Dashed white curves mark the contours where $\Delta \mathrm{RV}_{\rm max} = 20$\kms at different inclinations and $q=0.1$.}
   \label{fig:Pcrit}
\end{figure}

Given that our sample includes giant stars (luminosity class III) it is certainly possible that interactions might have already occurred. A lower surface gravity value, or analogously a larger radius, will affect the minimum orbital period before interaction ($P_{\rm crit}$) with consequences for the distribution of \drvmax and the multiplicity fraction \citep{badenes+18}. To quantify the effect of including moderately evolved stars, we compute $P_{\rm crit}$ \citep[following][]{mazzola+22} for a range of primary masses ($M_1=6\text{--}16$\Msun), surface gravities ($\log(g/{\rm cm\,s^{-2}})=3.0\text{--}4.5$), and mass ratios ($q=0.1\text{--}1$), representative of our sample of B-type stars. From $P_{\rm crit}$, we compute the allowed \drvmax values for a range of orbital inclinations between 20 and 90 degrees. Figure~\ref{fig:Pcrit} (top panel) shows how $P_{\rm crit}$ increases significantly towards lower $\log(g)$ values for a fixed mass ratio of $q=0.5$, illustrating that systems with orbital periods below 5\,d would have interacted once the primary reaches $\log(g)\approx3.5$. Using $q=1$ would shift the critical periods roughly by -0.2\,dex in $\log(g)$. In the bottom panel we show the corresponding maximum peak-to-peak RV amplitude (\drvmax$=2K_1$) for a typical inclination of $i=60^{\circ}$. Even at moderate inclinations and $q=0.5$, the minimum \drvmax remains around 100\kms for most of our parameter space, safely above the 20\kms threshold used in our binary criterion. Only in extreme cases of very low mass ratios ($q\lesssim0.1$) and low orbital inclinations ($i<40^{\circ}$) would \drvmax fall below 20\kms (dashed white contours). Therefore, even though the larger radii of giant stars shift the shortest detached periods to higher values, in most cases the resulting RV amplitudes remain well above our detection threshold.

\subsection{Observed multiplicity fraction}

\begin{figure}
   \centering
   \includegraphics[width=\hsize]{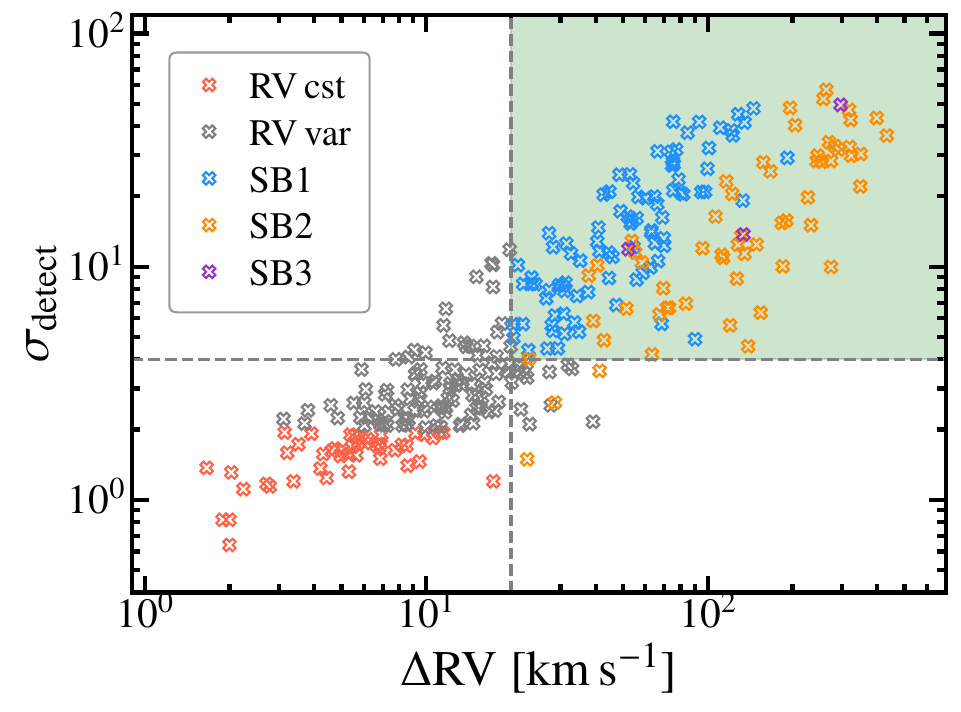}
      \caption{Spectroscopic binaries, RV\,var, and RV\,cst stars classified according to our binary criteria. Our adopted thresholds of $\sigma_{\rm detect}>4$ and $C>20$\kms are shown with dashed grey lines. Green region highlights systems fulfilling both criteria.}
      \label{fig:sigma_d}
   \end{figure}

According to the binary criteria, we have classified 153 stars as spectroscopic binaries, 91 SB1s, 59 SB2s, and 3 SB3s. To illustrate the distribution of our classifications based on the two criteria, we show a 2D diagram in Fig.~\ref{fig:sigma_d}. There are three SB2 systems that do not reach the $4\sigma$ significance threshold, these are BLOeM~2-023, BLOeM~3-020, and BLOeM~4-056. All these systems have been flagged as potential binary interaction products, and as discussed previously, the more massive star displays smaller RV variations and larger RV uncertainty. Because of their clear SB2 status we have included these systems in the determination of the multiplicity fraction. 

Here we defined the multiplicity fraction as that of the B-type primaries in our sample. Due to the unknown nature of most of the companions, we do not include them in our binary statistics. The multiplicity fraction quoted henceforth is effectively a close binary fraction that takes into account binaries at separations $a<10$\,au, or $\log(P_{\rm orb}/d)<3.5$, which roughly corresponds to the separations that will allow for binary interactions to happen. Consequently, this is the range of orbital periods for which observational biases are accounted for when determining the intrinsic binary fraction (see Sect.~\ref{sec:intrinsic_multp}). The 153 binaries in our sample yield an observed fraction of \fobs{50}{3}, where the uncertainty represents the standard error of a proportion based on binomial statistics, corresponding to a 68\% confidence interval. This is a significantly larger observed fraction than that found for the 30 Dor “unevolved'' stars \citep[$25\pm2\%$;][]{dunstall+15}. However, the detection probabilities of both surveys need to be carefully taken into account for a meaningful comparison. For example, the nine epochs analysed in this work provide an advantage over the six LR02 VFTS observations considered by \citeauthor{dunstall+15}. On the other hand, VFTS was much more sensitive to moderate-to-long period binaries due to its longer observational baseline of 10-12 months \citep[and up to 22 months for 31 targets;][]{dunstall+15}. We further discuss the differences between the samples of B-type stars and the implications for bias correction in Sect.~\ref{sec:intrinsic_multp}. 

It is, however, possible to compare the observed multiplicity of our sample with that of the \bloem O-type stars. \citet{sana+24} determined an observed binary fraction of \fobs{45}{4} for the 139 O-type stars in the BLOeM sample, which is in agreement with our findings for the B-type stars. This is interesting because the B-type population may suffer from different biases compared to the O-type stars due to their fainter magnitudes and lower S/N spectra. These biases result in a detection probability that drops at shorter orbital periods and larger companions masses for B-type stars \citep[see][and Sect.~\ref{sec:disc_bin_frac}]{sana+24}. 
A clear example of this is 30~Dor, where the observed multiplicity fraction was found to be $f^{\rm obs}_{\rm mult,\, O}=35.0\pm2.5\%$ for the O-type sample \citep{sana+13} and $f^{\rm obs}_{\rm mult,\, B}=25\pm2\%$ for the B-type stars \citep{dunstall+15}.
Given their faster evolution and higher detectability as binaries, the O-type population may be more affected by binary interactions. Binary interactions can effectively reduce the number of observable binaries through mergers or envelope stripping. Envelope stripping through stable mass transfer can lead to large mass ratios, making it difficult to detect RV variations and the flux contribution from the companion in optical wavelengths. Two possible scenarios might explain the similar observed multiplicity fractions of O- and B-type \bloem stars: (1) the initial multiplicity fraction of O-type stars is higher than that of B-type stars in the SMC but has decreased due to binary interactions; or (2) both populations have similar initial multiplicity fractions, but binary interactions are less frequent during the early phases of evolution at low metallicity. While the first option seems the most logical, there is also evidence that might support the latter possibility. As a result of the significantly reduced atmospheric opacity compared to stars in the LMC and MW, MS massive stars in the SMC are more compact, hotter, and more luminous \citep{maeder+meynet00, goetberg+17}, potentially delaying Roche-lobe overflow \citep{smith14}. Based on evolutionary models, \citet{klencki+20} found that at metallicities comparable to that of the SMC, the most common post-MS donors in massive binaries ({  $\sim$}16--50\Msun) are slowly expanding core-He burning (CHeB) stars. This means that interactions take place on average at later phases of evolution due to the smaller radii of low-metallicity massive stars. Even if this effect is not large enough to make a difference during the MS, it might play an important role for stars crossing the Hertzsprung gap, during which case-B mass transfer occurs. Although not the scope of this work, binary population synthesis modelling could help evaluate the significance of metallicity effects across different evolutionary phases and their impact on the multiplicity fraction of a population of massive stars, such as that of the \bloem OB-type sample.

\begin{figure}
   \centering
   \includegraphics[width=\hsize]{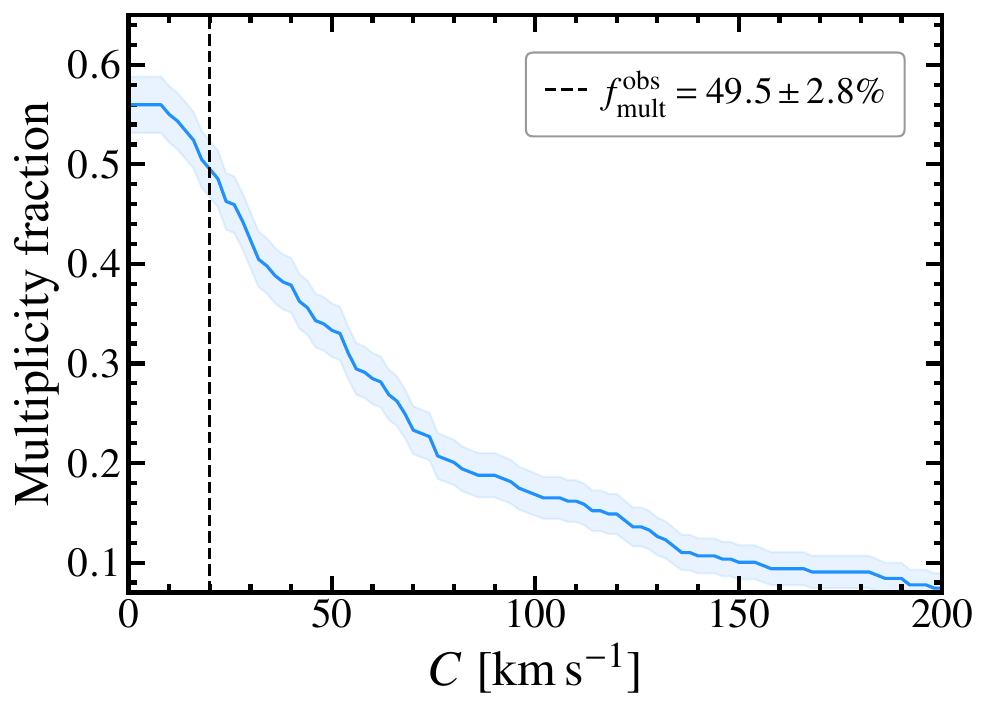}
      \caption{Observed multiplicity fraction as a function of the cut-off value $C$. The 20\kms threshold used by previous studies \citep{sana+13, bodensteiner+21, banyard+22} is marked with the dashed line. The adoption of $C=20$\kms yields the observed binary fraction of \fobs{49.5}{2.8}. At $C=0$, the observed multiplicity fraction is purely given by Equation~\ref{eq:sigmad}.}
      \label{fig:fbin}
\end{figure}

In Fig.~\ref{fig:fbin} we show the dependence of our observed multiplicity fraction on the adopted value of the $C$ cut-off. We do not observe a significant feature that might indicate a transition between variability caused by intrinsic features (e.g. winds or pulsations) and that caused by orbital motion. This is in contrast to what has been observed for the O-type stars and B-type supergiants in 30~Dor \citep{sana+13, dunstall+15}, where a distinct “kink'' is visible at 16\kms. It is not clear if this is a consequence of the less active atmospheres of the B-type stars at low metallicity or what the role of pulsations might be. However, since the winds of massive stars are much reduced at SMC metallicity \citep{vink+01, borklund+21}, we do not expect winds to play a major role in the RV variability of the B III/V stars \citep[e.g.][]{ramachandran+19}. Similarly, if pulsations were to strongly contribute to the RV variability, we would expect a feature as a kink close to 20\kms where the fraction of large RV amplitude variations drops significantly. We will come back to the role of pulsations in Sect~\ref{sec:disc_bin_frac}, where we discuss the impact of adopting different \drv threshold values.

\subsection{Variations in the observed multiplicity fraction}\label{ssec:fbin_var}

We have computed the observed binary fraction for each of the eight FLAMES fields in the BLOeM survey, as shown in Table~\ref{tab:field_fract}. We do not find a significant variation across fields, with most of them exhibiting multiplicity fractions close to 50\% within $1\sigma$. The exception is Field 2, which exhibits a multiplicity fraction of $74\pm10$\%. Compared to the mean fraction of the other seven fields (\fobs{50}{2}), this deviation appears statistically significant (${\rm Z{-}score}=2.4$ and ${\rm p}=0.02$). However, Field 2 has a relatively small sample size ($N{=}19$) compared to the other fields. The high uncertainty increases the likelihood that the observed difference is due to statistical fluctuations inherent in small sample sizes. We have not identified any environmental factors unique to Field 2 that would result in a higher multiplicity fraction. Therefore, we interpret this result cautiously and attribute the higher fraction in Field 2 to small number statistics rather than a genuine variation in the underlying stellar population.

\begin{table}
   \caption{Observed multiplicity fraction for each of the FLAMES fields.}
   \label{tab:field_fract}
   \center
   \begin{tabular}{cccc}
   \toprule
   \toprule
   Field & $N$ & Multiplicity Fraction \\
   \midrule
   1 &               24 &     0.54 ± 0.10 \\
   2 &               19 &     0.74 ± 0.10 \\
   3 &               20 &     0.55 ± 0.11 \\
   4 &               33 &     0.52 ± 0.09 \\
   5 &               42 &     0.50 ± 0.08 \\
   6 &               71 &     0.42 ± 0.06 \\
   7 &               47 &     0.43 ± 0.07 \\
   8 &               53 &     0.51 ± 0.07 \\
   \bottomrule
   \end{tabular}
   \tablefoot{$N$ is the number of early B-type stars in the field, and the quoted uncertainty represents the 68\% confidence interval.}
\end{table}

\begin{figure}
   \centering
   \includegraphics[width=\hsize]{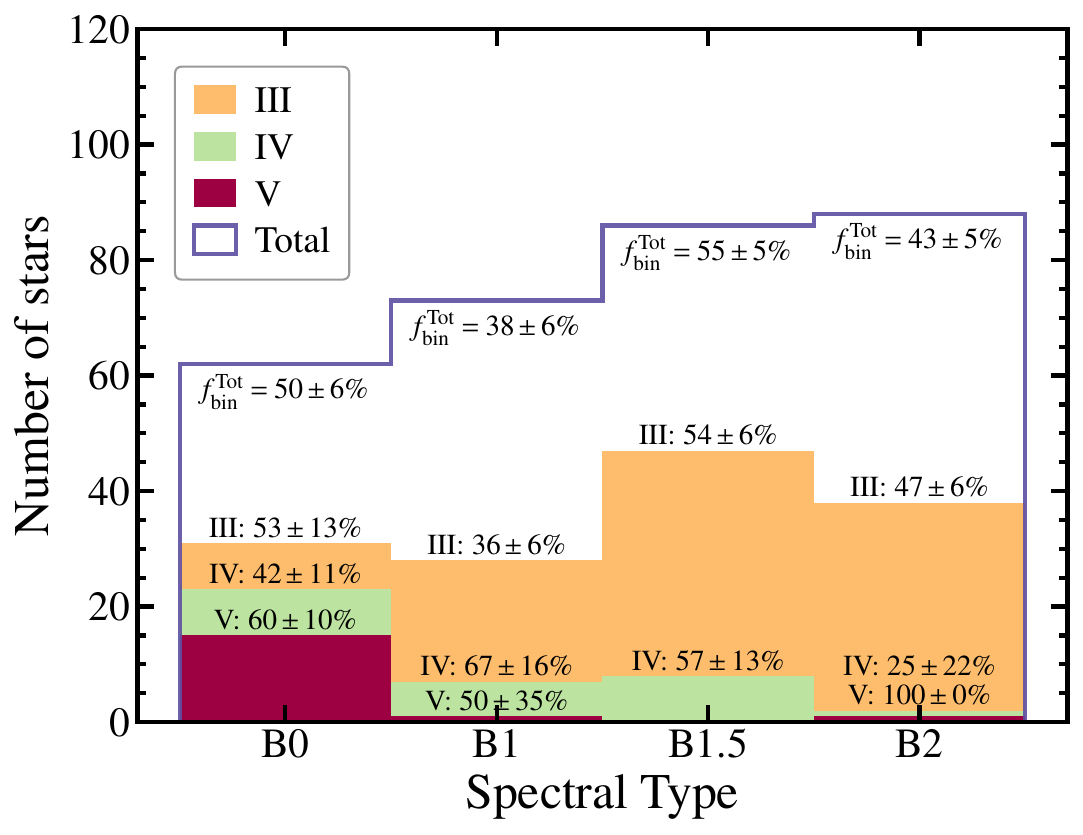}
      \caption{Histogram of our sample binned by spectral type. B0, B0.2, and B0.5 spectral types have been grouped in the B0 bin, as with B0.7 and B1 in the B1 bin, and B2 and B2.5 in the B2 bin, whereas B1.5 remains ungrouped. The unfilled histogram shows the total number of stars in each bin, whereas the filled histogram shows the number of identified binaries, stacked and colour-coded by luminosity class. Multiplicity fractions for each luminosity class have also been added as annotations, whereas the total multiplicity fraction is shown at the top.}
      \label{fig:fbin_SpT}
   \end{figure}

We have also investigated potential variations in the multiplicity fraction with spectral type, as a proxy for mass. In the histogram of Fig.~\ref{fig:fbin_SpT} we show the number of stars per spectral type and luminosity class, with their corresponding observed multiplicity fractions annotated. To minimise the impact of small number statistics, spectral types have been grouped as follows: the B0 bin includes spectral types from B0 to B0.5, B0.7 is grouped with B1, and B1.5 objects remain ungrouped, as in Fig.~\ref{fig:SpTs}. The few B2.5 stars have been incorporated into the B2 bin. The purple unfilled histogram represents the total number of stars in each spectral type bin, while $f^{\rm Tot}_{\rm bin}$ denotes the total multiplicity fraction within each bin, corresponding to the number of stars in the stacked histogram. Uncertainties in multiplicity fraction were computed using binomial statistics. Furthermore, the number of stars per luminosity class within each spectral type group is shown as stacked segments, each with its multiplicity fraction. It is not possible to draw strong conclusions from Fig.~\ref{fig:fbin_SpT} due to the small number of stars with luminosity classes V and IV. For the multiplicity fraction of giants and total fraction, there are no significant trends with temperature, suggesting that within our spectral range, the multiplicity fraction remains constant. 

\subsection{Intrinsic multiplicity fraction}\label{sec:intrinsic_multp}

To compute the intrinsic multiplicity fraction of our sample, we used a similar methodology to that applied in previous studies of OB-type samples \citep{sana+13, dunstall+15, bodensteiner+21, banyard+22} and other \bloem samples \citep{sana+24, patrick+25,britavskiy+25}.

We performed Monte Carlo (MC) simulations to estimate the detection probability of our observing campaign and correct for observational biases. We generated a synthetic population of stars where each star was randomly assigned to be either a single or a binary star based on a binomial distribution with a success probability $f_{\rm bin}$, representing the intrinsic multiplicity fraction \citep{sana+13}. For each binary system, we randomly drew orbital parameters, specifically orbital period ($P$), mass ratio ($q=M_2/M_1$), eccentricity ($e$), and primary mass ($M_1$), from assumed power-law distributions with specified indices and domains shown in Table~\ref{tab:distrib_ind}. 

Random orbital inclinations ($i$) and arguments of periastron ($\omega$) were assumed to account for random orientations in space. Each synthetic star was matched to a real star in our sample, inheriting its time sampling (observation time) and RV errors. Using the assigned orbital parameters and observation times, we computed the expected RV measurements for the synthetic binaries, incorporating the real RV errors to simulate observational noise. For single stars, we assumed constant RVs with noise corresponding to the measurement errors.

\begin{table}
   \caption{Power-law indices and range of our assumed distributions of orbital parameters for the computation of the detection probability and intrinsic multiplicity fraction.}
   \label{tab:distrib_ind}
   \center
   \begin{tabular}{lll}
   \toprule
   \toprule
   Parameter & Index & Domain \\
   \midrule
   $\log P$      & $\pi = 0 \pm 0.2$  &  [0, 3.5 ] \\
   $e$           & $\eta = -0.5 \pm 0.2$ & [0, 0.9] \\
   $q$           & $\kappa = 0 \pm 0.2$ & [0.1, 1] \\
   $M_1$\,[\Msun]  & $\gamma = -2.35$ & [8, 15] \\
   \bottomrule
   \end{tabular}
   \tablefoot{The primary mass ($M_1$) is obtained assuming a Salpeter initial mass function.}
\end{table}

We then applied our binary detection criteria described in Sect.~\ref{sec:crit} to the simulated observations to identify which synthetic binaries would be detected in our survey. By repeating this process 10\,000 times, we obtained robust statistics and estimated the detection probability (\pdet) as the fraction of synthetic binaries that met our detection criteria. This allowed us to compute the expected observed multiplicity fraction ($f^{\rm simul}_{\rm obs}$) given an intrinsic multiplicity fraction $f_{\rm bin}$.

   \begin{figure}
   \centering
   \includegraphics[width=\hsize]{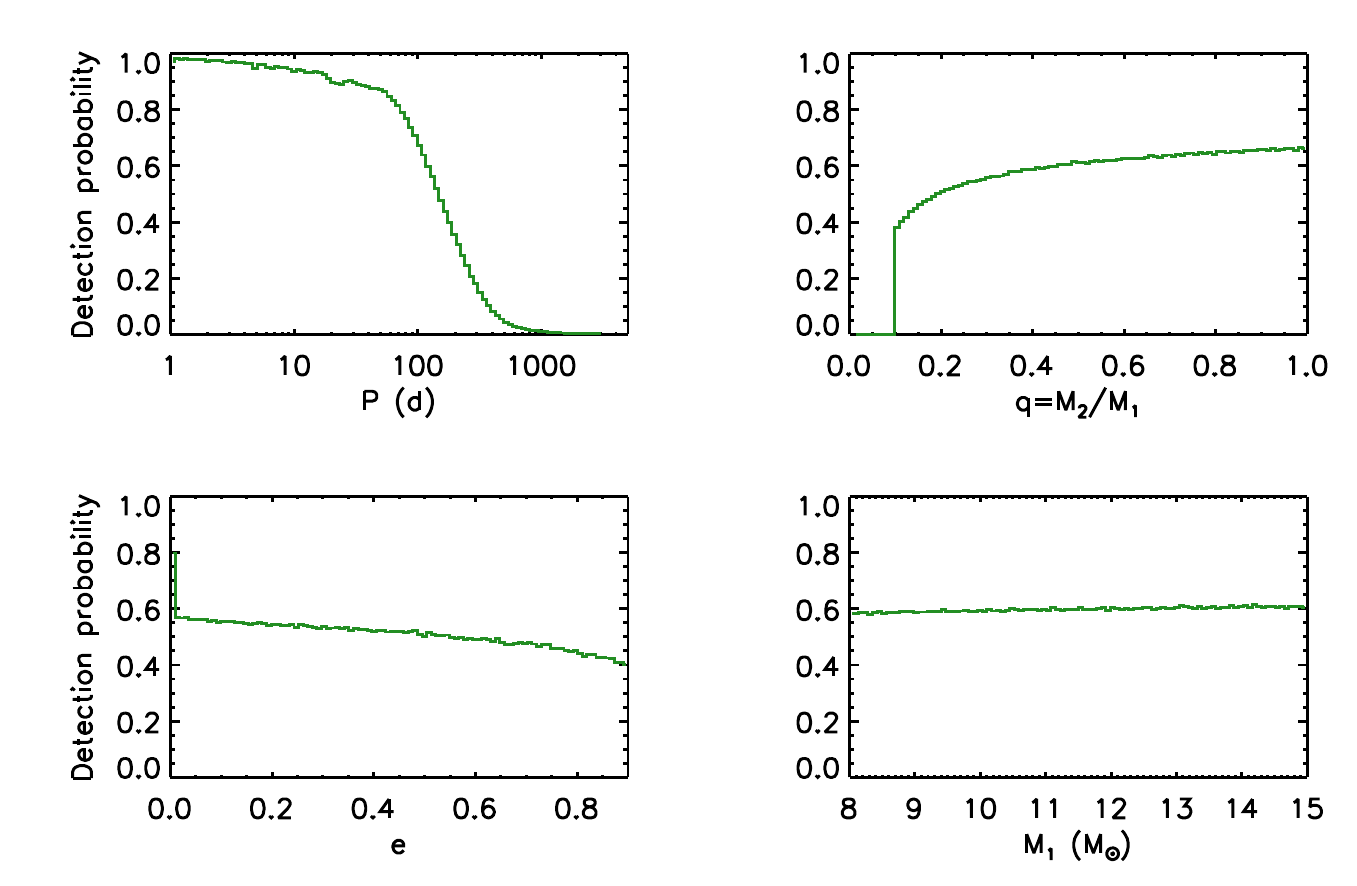}
      \caption{Detection probability of B-type binary systems in the BLOeM campaign, as a function of the orbital properties $P$, $q$, $e$, and $M_1$.}
      \label{fig:detect_prob}
   \end{figure}

The simulated observed multiplicity fractions are compared with the actual observed multiplicity fraction from our data. By adjusting $f_{\rm bin}$ in our simulations to match the observed fraction, we estimated the intrinsic multiplicity fractions of our sample. This approach accounts for observational biases and detection limitations of the \bloem campaign, providing a more accurate determination of the true multiplicity properties of our sample.

The detection probabilities for our sample of B-type binaries are shown in Fig.~\ref{fig:detect_prob}. As expected, we have a very high detection probability for short-period systems, $0.99^{+0.01}_{-0.02}$ for orbital periods up to 10\,d. After 60-70\,d the detection probability drops significantly,  reaching ${\sim}70\%$ for $P_{\rm orb}=100$\,d and ${\sim}35\%$ for $P_{\rm orb}=200$\,d, with an integrated probability for orbital periods between 10\,d and 1\,yr of $0.74\pm0.04$. When considering the full possible orbital periods range ($0<\log (P_{\rm orb}/{\rm d})<3.5$) the detection probability of the \bloem campaign for the B-type stars is \pdet$=0.62^{+0.07}_{-0.06}$. In the case of the mass ratio, the sharp drop is due to the minimum value considered (see Table~\ref{tab:distrib_ind}). None of $q$, $e$, or $M_1$ have as much effect in the detection probability as $P$.

After accounting for the observational biases in the \bloem campaign and given our detection probability, we found an intrinsic multiplicity fraction of $f_{\rm mult}=80\pm8\%$. We can compare this value to the intrinsic multiplicity fraction found for the B-type stars in 30~Dor of $f_{\rm mult} = 58\pm11\%$ \citep{dunstall+15}. Performing a Z-test, we found these two multiplicity fractions to deviate by $3.2\sigma$, which is statistically significant at the 1\% $\alpha$--level (p=0.0015). The higher SMC fraction is even more relevant if we consider that the 30~Dor sample was dominated by luminosity class V stars (close to 60\%) whereas our III/V sample contains a majority of giant stars, which as seen in Fig.~\ref{fig:Pcrit}, cannot avoid interactions below orbital periods of a few days. The difference is even larger ($4.9\sigma$) when compared to $f_{\rm mult}=52\pm8\%$ of the galactic B-type stars in NGC~6231 \citep{banyard+22}, and to $f_{\rm mult}\approx55\%$ ($3.2\sigma$, assuming a mean uncertainty of 10) found for the OB-type stars in the Cygnus OB2 association \citep{kobulnicky+14}. This suggests a trend with metallicity that will be explored in more detail in Sect.~\ref{sec:disc_bin_frac}. 

   \begin{figure}
   \centering
   \includegraphics[width=\hsize]{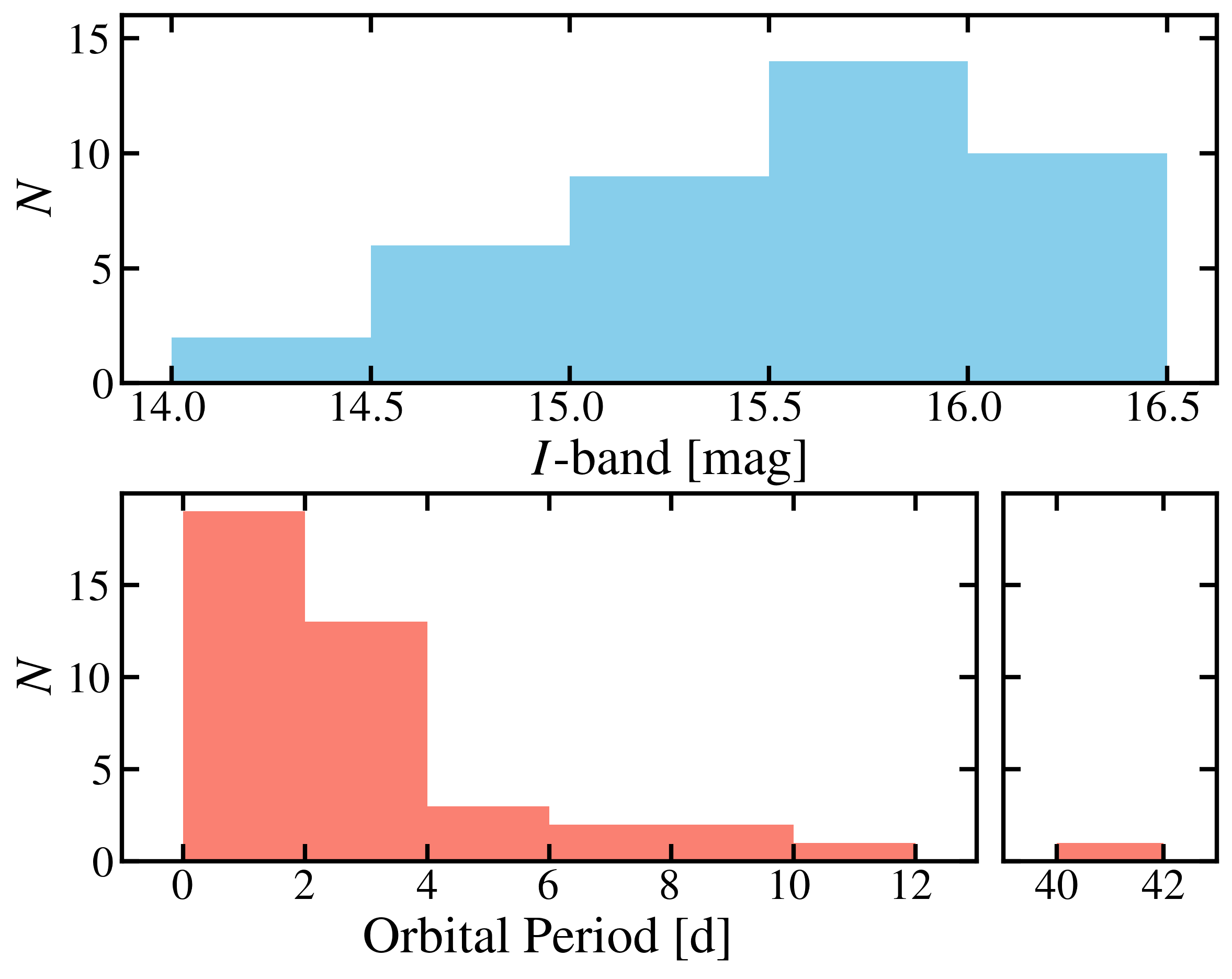}
      \caption{Distribution of $I$-band magnitudes (top) and photometric orbital periods (bottom) of the systems in our sample flagged as EBs by OGLE \citep{pawlak+16}.}
      \label{fig:ogle_periods}
   \end{figure}

\section{Determination of orbital periods}\label{sec:Porb}

\subsection{Eclipsing binaries}\label{sec:EB}

Some of our targets have been identified as EBs by OGLE \citep{pawlak+16}. We found 41 EBs which are listed in Table~D.2, including their photometric periods, $I$-band magnitudes, and light curve type. Their distribution of $I$-band magnitudes and orbital periods are shown in Fig.~\ref{fig:ogle_periods}. The 41 EBs correspond to 27\% of our binary sample. Among them, there are ten SB1 systems, 26 SB2s, all three SB3s, and two \rvvar stars. The latter two systems did not meet our binary criteria, but we found the correct orbital period for one (BLOeM~6-061), and the second system (BLOeM~7-054) has an orbital period just outside our frequency grid (1.099\,d, see Sect.~\ref{sec:bloem_periods}).	

Most of the EBs have been classified as detached or semi-detached by \citet{pawlak+16}, but there are two ellipsoidals and one contact system. Both ellipsoidal systems are SB2s that suffered from misidentification of their components. However, we still recovered half of the photometric orbital period in both cases, as is common for ellipsoidal variables. The only contact system, BLOeM~6-010, also suffered from component misidentification and it was flagged as a possible SB3, but we found the correct orbital period of 1.21\,d. Despite finding the same period (or half in some cases) as the OGLE period for these systems, the significance in all three cases was too low to confidently list them as the correct periods (see details below).

We have done a preliminary inspection of the photometric data of the EBs mainly to find evidence and confirm third components. As mentioned in Sect.~\ref{sec:crit}, we found clear signs of third companions in three systems that we have classified as SB3, but also evidence of potential contributions from further components in another 15 systems. Our photometric analysis indicates that there are at least eight systems where there is likely a third companion, which would represent a lower limit of 3\% of triple systems in our sample. We provide comments on these and other systems in Appendix~\ref{append:indiv_sys}.

\subsection{Period search in the \bloem data} \label{sec:bloem_periods}

We have applied the Lomb-Scargle (LS) periodogram \citep{lomb76,scargle82}, as implemented in \citet{villasenor+21}, to our sample to search for periodicity in our RVs. Given the baselines for the different FLAMES fields, we could be sensitive to periods up to ${\sim}$40--60\,d. The nine epochs of observations (less in some cases, see Sect.~\ref{sec:sample}) allowed us to obtain significant signals for 45 systems, which corresponds to 29\% of the binaries.

   \begin{figure}
   \centering
   \includegraphics[width=\hsize]{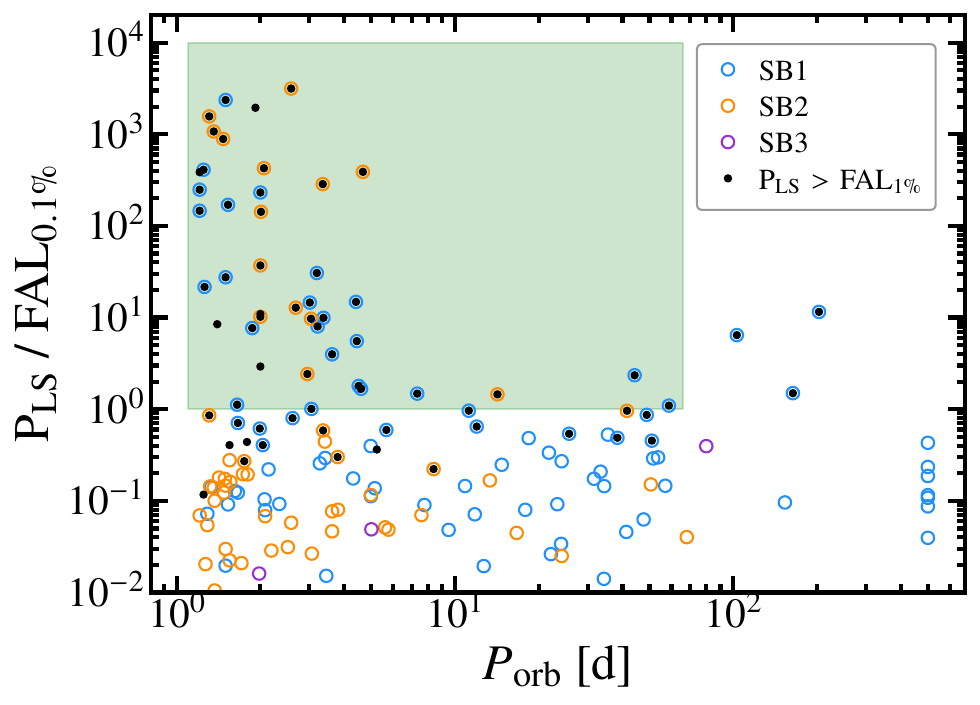}
      \caption{Significance of the LS periodogram peaks as a function of the recovered periods. We consider a period significant if its peak surpasses a FAL of 0.1\%. Green rectangle shows the significant periods according to our criterion and the cadence and baseline of the \bloem campaign.}
      \label{fig:LS_pow}
   \end{figure}

The significance of the signal has been defined with respect to the false-alarm-probability (FAP). All peaks above 1.1\,d of period in the LS periodogram with a LS power (P$_{\rm LS}$) above a false-alarm level (FAL) of 0.1\% (i.e. FAP below 0.1\%) are considered significant. It is possible that periods with a FAP above this threshold might be correct, but the signal is not strong enough due to the number of epochs, the intrinsic properties of the system (such as high eccentricity), or difficulties in the fitting procedure (e.g. component misidentification). To illustrate our criterion, we show in Figure~\ref{fig:LS_pow} the P$_{\rm LS}$ of the highest peak in the LS periodogram (above 1.1\,d) over the power at a FAL of 0.1\% as a function of the orbital period. Systems are colour coded by binary classification (open circles), and we highlight systems with P$_{\rm LS}$ above a FAL of 1\% with black filled circles. The green region encloses systems with orbital periods considered as significant. There are three systems with periods larger than 100\,d; from the visual inspection of their RVs, they do present a variation consistent with long orbital periods, but we do not have enough coverage of the orbit to compute an accurate period.

\subsection{Comparison with 30~Dor B-type binaries}

\subsubsection{Orbital periods}

The B-type Binaries Characterisation programme \citep[BBC;][]{villasenor+21} studied 88 of the binary candidates identified by \citet{dunstall+15} with 29 FLAMES observations over a baseline of 427\,d. \citet{villasenor+21} found orbital solutions for 50 SB1 and 14 SB2 systems, and potential solutions for 20 more systems. We can compare the distributions of orbital periods for both populations, over a similar baseline. For this, we have selected nine epochs from the BBC campaign covering a baseline of ${\sim}$60\,d, without including the low S/N epochs 10, 12, 13, 17 (see their Table 1 and data reduction section). We computed the LS periodogram for the corresponding RVs for all systems in the BBC sample. 

   \begin{figure}
   \centering
   \includegraphics[width=\hsize]{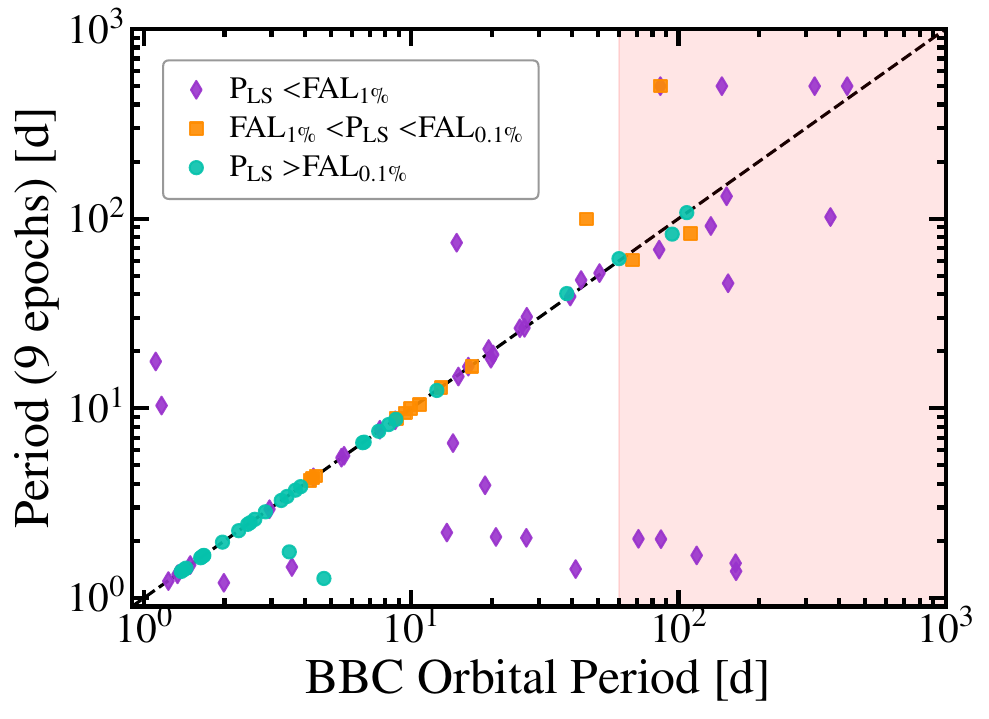}
      \caption{Comparison of BBC orbital periods \citep{villasenor+21} with periods derived from nine epochs. The dashed diagonal line represents equal periods. Data points are categorised based on their LS Power (P$_{\rm LS}$) relative to the power at a false alarm level (FAL). The shaded red area highlights period above 60 days.}
      \label{fig:bbc_P}
   \end{figure}

The results can be seen in Fig.~\ref{fig:bbc_P}. Similarly to the periods for the \bloem sample, we distinguish between those systems whose P$_{\rm LS}$ exceeds the 0.1\% FAL (turquoise circles, 29 system), those between the 0.1 and 1\% thresholds (orange rectangles, 13 systems), and those with P$_{\rm LS}$ below the 1\% threshold (purple diamonds, 45 systems). We found excellent agreement for the first two groups. When limiting the results to a maximum period of 60\,d, we found 93\% agreement for systems with P$_{\rm LS}$ above the 0.1\% threshold, and 90\% agreement for those between 0.1--1\%. For this reason we use the periods from both these groups in the comparison with the \bloem sample, resulting in a sample of 34 systems.

In the case of the \bloem periods, our test with the BBC sample reaffirms our confidence in the periods that met our significance criterion. However, we restrict the periods to those highlighted in Fig.~\ref{fig:LS_pow}, not including those with P$_{\rm LS}<0.1\%$\,FAL. We do include the orbital periods from OGLE, for a total of 57 objects.

\begin{figure}
   \centering
   \includegraphics[width=\hsize]{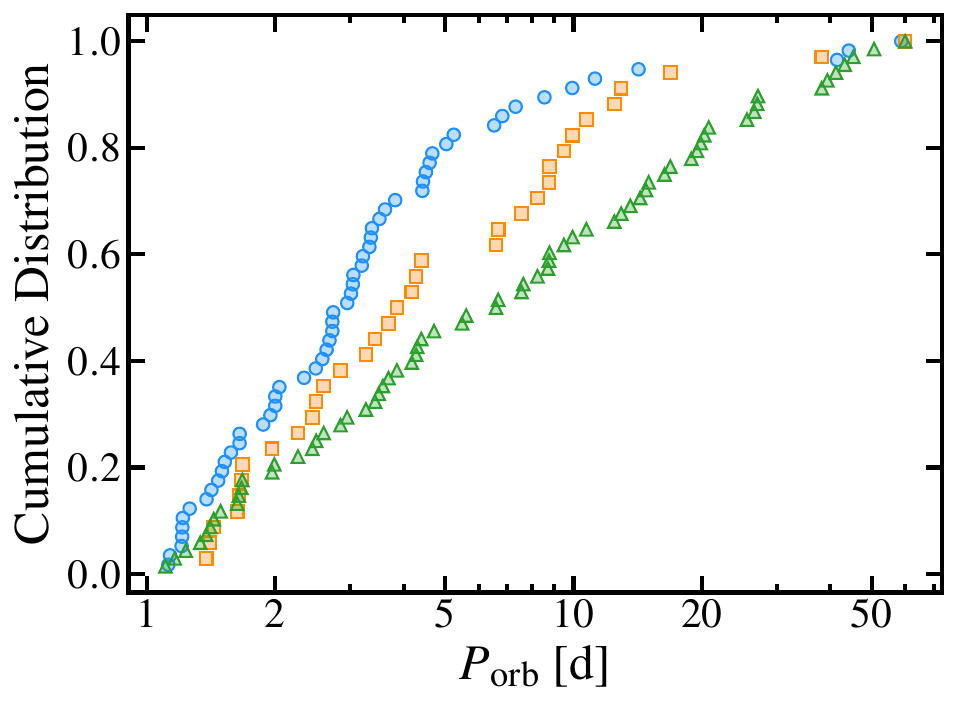}
      \caption{Cumulative distribution of the orbital periods of the \bloem B-type binaries (blue circles, 57 systems) compared to those of the LMC sample from the BBC programme \citep{villasenor+21} in orange rectangles (nine epochs, 34 systems) and green triangles (all epochs, 68 systems).}
      \label{fig:cum_dis}
   \end{figure}

The cumulative distribution of orbital periods is shown in Fig.~\ref{fig:cum_dis}. For comparison, we also plotted the distribution of periods for all BBC systems with a period below 60\,d (green triangles). When comparing both samples with nine epochs, there is only a marginal difference around 6\,d. To assess the significance of this difference, we have computed the Kolmogorov-Smirnov (KS) test, obtaining a KS-statistic of $D=0.24$ and a p-value of 0.16, which does not reach significance at the 10\% level. However, when comparing with the full BBC sample up to periods of 60\,d, we do find significant differences ($D=0.37$, p=0.0003). It is clear that both samples with nine epochs lack systems with periods larger than ${\sim}15$\,d in contrast to the full BBC sample which approaches a uniform distribution in $\log P$. Judging from Fig.~\ref{fig:bbc_P}, the significance of the periods found with the LS periodogram starts to decrease after 10\,d, and from that point up to 60\,d we see a majority of systems with P$_{\rm LS}< 1\%$\,FAL, even though we did find the correct periods for many of them. 

Once all the 25 epochs of BLOeM observations are assembled, we will be able to recover accurate orbital solutions for most of the binaries and compare with the BBC and other samples of B-type stars over the full range of orbital periods.

\subsubsection{Distribution of \drvmax}

\begin{figure}
    \centering
    \includegraphics[width=\linewidth]{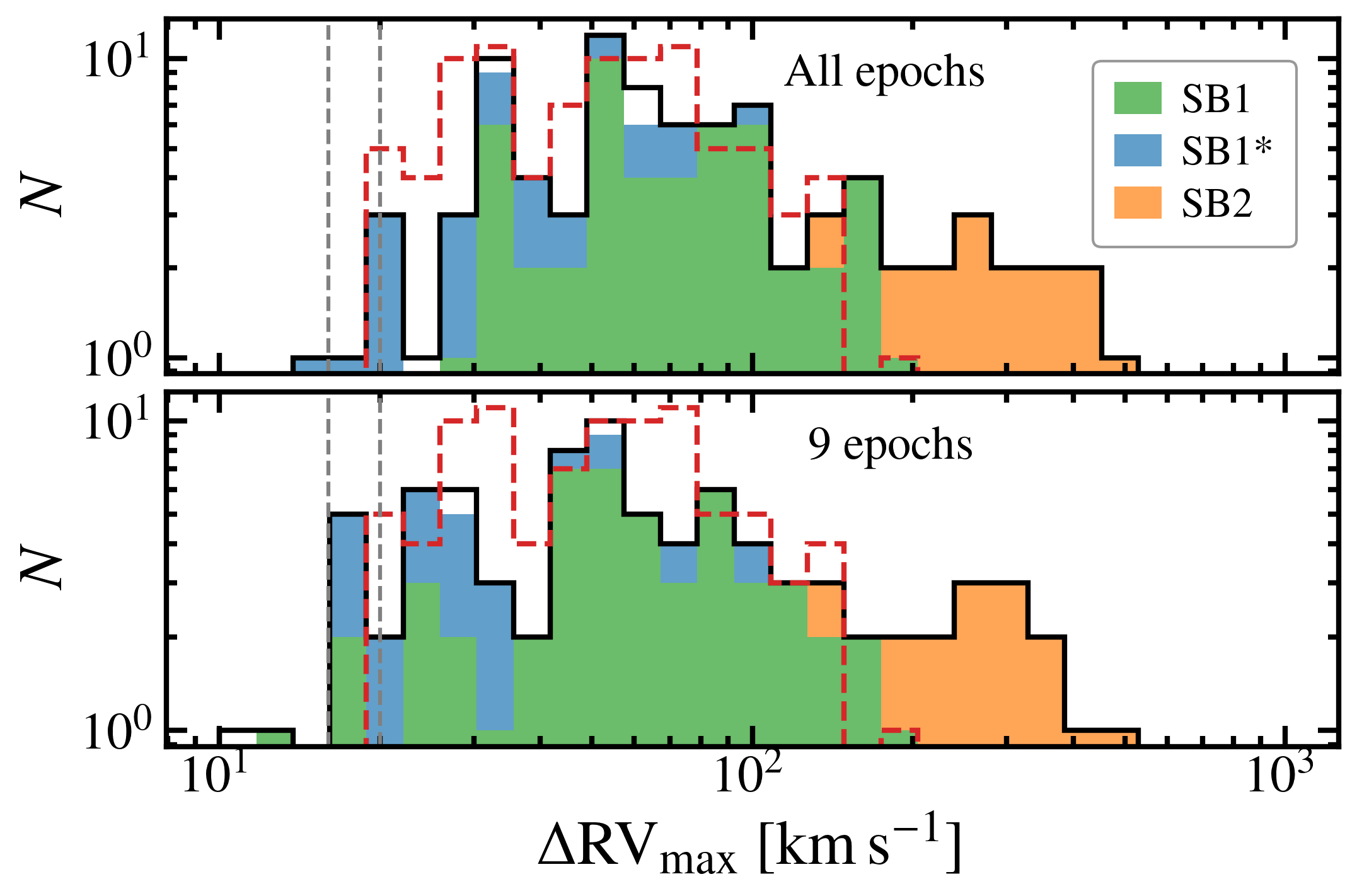}
    \caption{Distributions of \drvmax for the BBC sample. Top panel: \drvmax derived from epochs used for period determination in BBC ($\sim$25 per star). The different binary classifications from \citet{villasenor+21} are shown in the stacked histograms. SB1* systems were defined as those with less robust orbital periods, whereas those without orbital solutions are shown in white. Bottom panel: same as above but using only nine randomly selected epochs per star. For comparison, we include the distribution of \bloem SB1 systems in the dashed red histogram. The dashed vertical lines at 20 and 16\kms highlight threshold values discussed in Sect.~\ref{sec:crit}.}
    \label{fig:drv_bbc}
\end{figure}

We also examined the distribution of \drvmax values for the BBC systems with determined periods. Given the larger number of epochs (29) of the BBC programme and the reliable periods, their \drvmax values are robust. The distribution is shown in Fig.~\ref{fig:drv_bbc}, where we have colour-coded the stacked histogram by the binary classifications used by \citet{villasenor+21}. The top panel shows the \drvmax values computed from all the epochs with well determined RV measurements that were used in the computation of orbital solutions. In the bottom panel, we plot the distribution obtained by only considering nine epochs as in the case of the \bloem sample. Comparing the two panels, we see that the distribution is shifted slightly towards lower values when only nine epochs are used, with long-period, eccentric binaries being the most affected. Indeed, the three SB1 systems that fall below 20\kms all have $P_{\rm orb}>150$\,d. Notably, 34 of the BBC SB1 systems have \drvmax values below 80\kms, using a large binary threshold such as this (see discussion in Sect.~\ref{ssec:RVthresh}) would miss 64\% of the SB1 systems. On the other side, a threshold of 20\kms would still skip some of the confirmed long-period binaries in BBC when only using nine epochs, whereas only three systems without orbital solutions remain above the threshold. This evidence also suggests that 20\kms is an appropriate limit to separate binarity from intrinsic variability.

\section{Discussion}\label{sec:disc}

\subsection{Trend with metallicity}\label{sec:disc_bin_frac}

\begin{figure}
   \centering
   \includegraphics[width=\hsize]{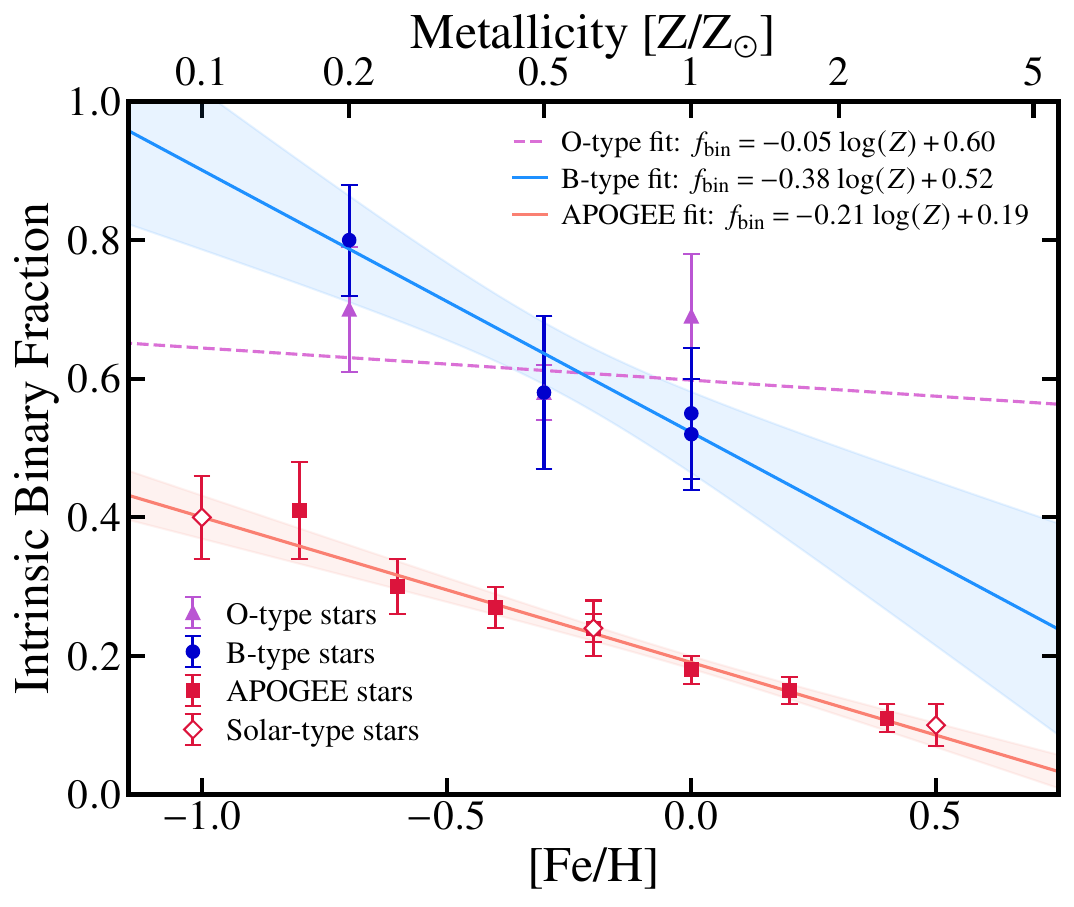}
      \caption{Dependence of the multiplicity fraction of different mass regimes on metallicity. The B-type studies at different metallicities are represented by blue circles; $Z_{\rm SMC}=0.2$\Zsun (this work), $Z_{\rm LMC}=0.5$\,\Zsun \citep{dunstall+15}, $Z_{\rm MW}=$\Zsun \citep{kobulnicky+14,banyard+22}. The best fit to the data with the 68\% confidence interval is represented by the blue line and shaded area. Red squares show the sample of G- and K-type dwarfs from \citet{badenes+18}, whereas red diamonds correspond to the combined data of five solar-type samples presented by \citet{moe+19}, see details in main text. The dashed purple line is the linear regression found by \citet{sana+24} 
      for the \bloem O-type stars (purple triangles), included for comparison.}
      \label{fig:metallicity_fbin}
   \end{figure}

As seen in Sect.~\ref{sec:intrinsic_multp}, the large multiplicity fraction we have found for the \bloem B-type stars suggests a trend with metallicity. To test this dependence, we performed a simple linear regression $f_{\rm mult}=b+m\,\log(Z)$ (as the one applied by \citealt{sana+24} to the O-type stars) to the results from \citet{kobulnicky+14,dunstall+15,banyard+22}, and our result, where we have adopted as uncertainty for the former the mean of the errors of the other studies, and for simplicity, in the case of asymmetric errors, the mean between them. The fit is shown in Fig.~\ref{fig:metallicity_fbin}, with its 68\% confidence interval in shaded blue. We found a slope of $m=-0.38 \pm 0.14$, which is $2.6\sigma$ away from zero. 

To illustrate and compare with the case of solar-type stars, we include in the plot a sample of G- and K-type IV/V stars observed by APOGEE and first presented by \citet{badenes+18}. Close binary fraction values as a function of metallicity were obtained from \citet[][their figure 11]{moe+19}, with the corresponding fit shown in red. We also include three additional values (red diamonds, not included in the linear regression) at [Fe/H]$\,=\,$$-$1.0, $-$0.2, and $+$0.5. These points represent a weighted moving average from five different solar-type samples, including the \citeauthor{badenes+18} sample, analysed by \citet[][see their figure 18]{moe+19}. The averaged values are in good agreement with the former measurements, but we only fit the APOGEE sample since it is much larger ($\sim20\,000$ objects) and presents the lowest errors. For the details of the solar-type samples we refer the reader to \citet{moe+19}, but we note that all samples were corrected for incompleteness up to orbital periods of $\log(P/{\rm d})=4$. We have computed the linear fit to the APOGEE data and found a slope of $-0.21 \pm 0.03$. 
The two slope values for solar and B-type stars are not significantly different from each other when compared with a t-test ($p=0.36$), suggesting a similar trend of the multiplicity fraction with metallicity. Figure~\ref{fig:metallicity_fbin} also shows the linear fit found by \citet{sana+24} for the O-type stars (dashed purple line) with its corresponding MW \citep{sana+12}, LMC \citep{sana+13, almeida+17}, and SMC \citep{sana+24} multiplicity fractions. No clear relation was found for the O-type stars with metallicity by \citet{sana+24}\footnote{For a comparison of binary fractions with the O-type stars and the other \bloem samples see Appendix~\ref{append:HRD}}.

To assess the significance of the trend between metallicity and multiplicity fraction for the B-type stars, we conducted a t-test under the null hypothesis that the slope of the linear regression is consistent with zero. The analysis yielded a p-value of $p=0.12$, meaning that we cannot reject the null hypothesis at the conventional significance level of 0.05. This indicates that the observed trend is not statistically significant. The lack of significance may be attributed to the small sample size, which limits the statistical power of the test and increases the uncertainty in the slope estimate. With only four observations, the degrees of freedom are low, resulting in a t-distribution with heavy tails. This makes it more challenging to achieve statistical significance, even when the estimated slope suggests a potential trend. As an example, we repeated the test on the APOGEE and the solar-type averaged samples, with the latter returning $p=0.13$ from the three measurements, whereas the APOGEE sample with seven measurements (and smaller errors) returned $p=0.008$. Future studies at different metallicities are necessary to more reliably determine if the trend with metallicity for the B-type stars is a real feature.

The increased multiplicity fraction found for the SMC B-type stars in this work, contrasts with the case of EBs. \citet{moe+distefano13} found multiplicity properties of EBs with early B-type primaries to be invariant with metallicity in the range $-0.7 < \log \left(Z/{\rm Z_{\odot}}\right) < 0.0$. Recently, \citet{menon+24} estimated the multiplicity fraction across EBs in the LMC and SMC to also be roughly constant. However, with EBs it is only possible to probe the close-binary regime due to the limited range of orbital inclinations that would lead to eclipses, which decreases with separation. Indeed, the sample from \citet{moe+distefano13} had orbital periods up to 20\,d. This can also be seen in Fig.~\ref{fig:ogle_periods} where we can see only two EBs in our sample with periods above 10\,d, and with most of the EBs having periods of less than 4\,d. To reconcile the multiplicity fraction of short-period EBs with our increased multiplicity fraction at low $Z$, some type of variation in the formation mechanism of massive binaries would be needed. 

Close massive binaries ($a{\,<\,}10$\,au) are believed to form through disc fragmentation followed by inward orbital migration via circumbinary accretion \citep{tokovinin+moe20}. Those in the closest orbits ($P_{\rm orb}<10$\,d) require extremely early fragmentation, coupled with subsequent envelope mass accretion to harden the orbits to such short periods, a process that remains largely independent of metallicity. This would explain why EBs, which predominantly probe these very close binaries, exhibit multiplicity properties that are invariant with metallicity. This is similar to what we see for the more massive O-type stars. In order to form an O-type star, a high accretion rate and large disc surface density are needed, which do not depend on metallicity. This presents a natural explanation for why the multiplicity fraction of O-type stars remains high at different metallicities.

However, for B-type binaries at wider separations ($0.2<a<10$\,au), metallicity might have played a more significant role. Metal-poor accretion discs, such as those in the SMC, are less optically thick and can cool more efficiently. This enhanced cooling and reduced opacity facilitate gravitational instabilities, leading to increased disc fragmentation. Therefore, for stars where fragmentation occurs at later times, there is not enough remaining disc mass or accretion to dissipate the orbital energy effectively and decrease separation beyond $a<0.2$\,au. This mechanism would potentially result in a larger multiplicity fraction in early B-type binaries with separations between 0.2 and 10\,au at low metallicity.

\subsection{Considering a higher RV threshold}\label{ssec:RVthresh}

The discussion so far has assumed that the \drv threshold of $C=20$\kms has no effect on our binary fraction. Indeed, by correcting by the observational biases, we are taking into account the binaries that did not pass our detection criteria. The recovered fraction of binaries should be independent of the choice of the threshold, otherwise it is possible that we are heavily contaminated by single stars with intrinsic variability. To test this hypothesis, we have repeated our analysis adopting values of $C=50$\kms and $C=80$\kms and found an anti-correlation with the multiplicity fraction. For $C=50$\kms, the intrinsic multiplicity fraction is reduced to $f_{\rm mult}=70^{+12}_{-9}\%$, whereas for $C=80$\kms it further drops to $f_{\rm mult}=56^{+11}_{-8}\%$.  

The lower intrinsic multiplicity fraction has clear consequences for the interpretation of our analysis. Perhaps the most relevant one is the effect on the metallicity trend.  
This lower value argues in favour of a constant multiplicity fraction across the metallicities of the MW, LMC, and SMC, very similar to the one found for the \bloem O-type stars. The metallicity of the SMC is representative of galaxies at redshifts $z=3$ \citep{sommariva+12} and it can be found at redshifts as high as $z=10$ \citep{nakajima+23}, suggesting an universality in the formation process of O- and B-type binaries. However, we note that the RV threshold here is only applied to the \bloem sample, it would be interesting to see if the effect on the multiplicity fraction is also observed for the other samples of OB-type stars, but this is beyond the scope of our work.

However, it is crucial to understand where the dependence on the \drv threshold comes from if we want to draw any conclusions. One possible explanation for the variation of the intrinsic binary fraction is that there are other sources of variability in play contaminating our binary sample, different from binary motion. Indeed, our sample intersects the predicted instability region of $\beta$~Cephei pulsating stars, which span the mass range of about 8 to 25 \Msun, extending from the MS to the giant phase \citep{Bowman2020}. However, the bulk of RV amplitudes caused by pulsations concentrates in the 10-20\kms range, with a tail that can go up to 40--50\kms \citep{stankov+handler05, hey+aerts24}, and periods roughly between 2 and 12\,h \citep{aerts10}. 

Even if we are heavily contaminated by $\beta$~Cephei pulsators, it is hard to explain the drop in multiplicity fraction when increasing $C$ from 50 to 80\kms. We do not expect pulsation to play a significance role at that scale of variability. Furthermore, the mechanism causing pulsations is traditionally explained as variations in the star's opacity driven by elements of the iron group \citep{aerts10}. Because the presence of significant iron is needed for the opacity enhancement, pulsating massive stars are rare in low metallicity environments, as in the SMC \citep{Bowman2020}. Therefore, we would at least expect the role of pulsators to be smaller in comparison to the MW and LMC.

We are not aware of another source of RV variability that could play such a significant role at these high \drv values ($\geq50$\kms). With the full number of epochs available, robust orbital periods will allow us to more effectively disentangle the binary and pulsation contribution, since we are not sensitive to the periods produced by pulsations in $\beta$~Cephei stars. Moreover, time series photometry may help elucidate (high-amplitude) pulsators among the BLOeM sample and the wider population of SMC massive stars \citep[see][]{bowman+24}.

An alternative explanation is that our assumed distributions of orbital parameters are incorrect. Our results are particularly sensitive to the orbital period distribution, for which we have assumed an {\"O}pik law (see Table~\ref{tab:distrib_ind}). As discussed in Sect.~\ref{sec:Porb}, the fraction of binary system with periods between ${\sim}3$--10\,d seems to be larger for the \bloem sample than for the BBC sample of binaries. Although our analysis of the orbital period distribution is inconclusive about the significance of the difference in the distribution of both samples according to the KS test, there are other pieces of supporting evidence that are worth examining. First, the fraction of SB2 systems is much larger among the \bloem B-type binaries in comparison to the 30~Dor counterpart, with 39\% of the binaries (41\% if we include the SB3 systems) versus only 17\% in the BBC sample. There are several factors influencing the detection of SB2 systems, some are intrinsic to the system such as orbital period, mass ratio, and rotational broadening, but external factors like S/N can also play a role. \citet{villasenor+21} estimated a mean S/N for the BBC sample of 33. In contrast, the \bloem sample has a mean S/N of 55, which is significantly higher, probably responding to the higher fraction of giant stars. Moreover, the fraction of SB2s is similar to that of the VFTS O-type binaries \citep{almeida+17}, which presented similar S/N to our sample. In conclusion, the fraction of SB2 systems, although large, is not impossible to explain.

Eclipsing binaries are also abundant among the \bloem B-type III/V sample, reaching 27\% of the binaries, compared to only 12\% found by the BBC programme \citep{villasenor+21}. This could also be seen as evidence of an overabundant fraction of short-period systems. However, contrarily to BBC, our \bloem sample is dominated by giants which are roughly a factor two larger than MS stars. Since the probability of eclipses scales with radius, giant primaries will have a higher probability of being seen as EBs. Only accounting for MS stars, we find that similar fractions of the BBC and \bloem binaries are MS EBs (14\% and 11\% respectively).

We further test the possibility of deviating from an {\"O}pik law with Markov chain Monte Carlo (MCMC) simulations in Sect.~\ref{sec:porb_simulations}.

\begin{figure}
    \centering
    \includegraphics[width=\linewidth]{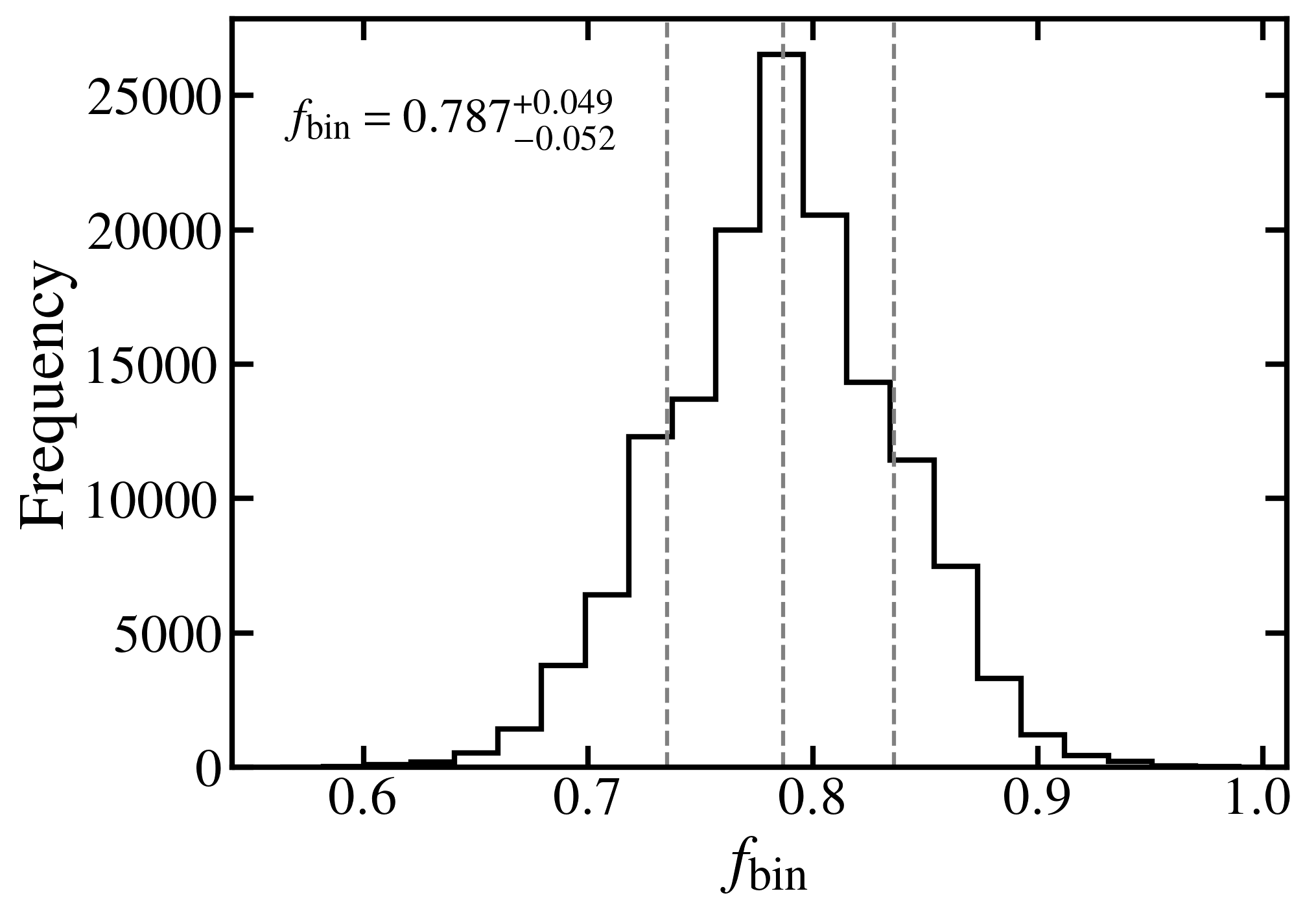}
    \caption{Posterior distribution for the binary fraction, $f_{\rm bin}$, from our MCMC simulations. Dashed grey lines indicate the 16th and 84th percentile.}
    \label{fig:posterior_fbin}
\end{figure}

\subsection{Binary fraction from the distribution of $\Delta{\rm RV}_{\rm max}$}

Even though we cannot explain the decrease in the intrinsic multiplicity fraction with the assumed RV threshold, we can explore other methods to obtain the multiplicity fraction. The distribution of \drvmax has been used to determine the binary fraction of various stellar populations, including white dwarfs \citep{badenes+maoz12, maoz+12} and low-mass stars \citep{badenes+18, mazzola+20}. These studies show that the distribution is characterised by a ``core'' of low \drvmax values, dominated by single stars and RV uncertainties, and a ``tail'' of higher values associated with close binaries. By reproducing these signatures with simulations, one can infer the intrinsic binary fraction of a sample without relying on explicit binary criteria.

We adopt an MC method similar to that described in Sect.~\ref{sec:intrinsic_multp}, sampling the same physical and orbital parameters but extending the orbital period range down to $\log P=-0.15$ to include sub-day periods observed by OGLE. We also extend the mass-ratio distribution to $q=0.01$ to probe lower-mass companions. We assumed circular orbits ($e=0$) for systems with orbital periods below 2\,d \citep{moe+distefano17}, whereas for longer orbits we sample $e$ from 0 to $e_{\max}$ (as given by \citet{badenes+18}). Values of the distribution indices remain as in Table~\ref{tab:distrib_ind}. To reduce stochastic noise, we generate 10\,000 stars with a binary fraction given by $f_{\rm bin}$ and use a MCMC approach to fit the observed \drvmax distribution by varying $f_{\rm bin}$. We scaled each simulated \drvmax distribution by the ratio of observed to simulated stars, and then compare with the real data using a Poisson-based log-likelihood \citep{cash79} in log-space intervals:

\begin{equation}
    \log \mathcal{L} \;=\; \sum_i \Bigl[\,n_i^\mathrm{obs} \,\ln(n_i^\mathrm{mock}) \;-\; n_i^\mathrm{mock}\Bigr],
\end{equation}
where $n_i^\mathrm{obs}$ is the number of observed stars in bin $i$ and $n_i^\mathrm{mock}$ is the corresponding (scaled) count of simulated stars.

\begin{figure}
    \centering
    \includegraphics[width=\linewidth]{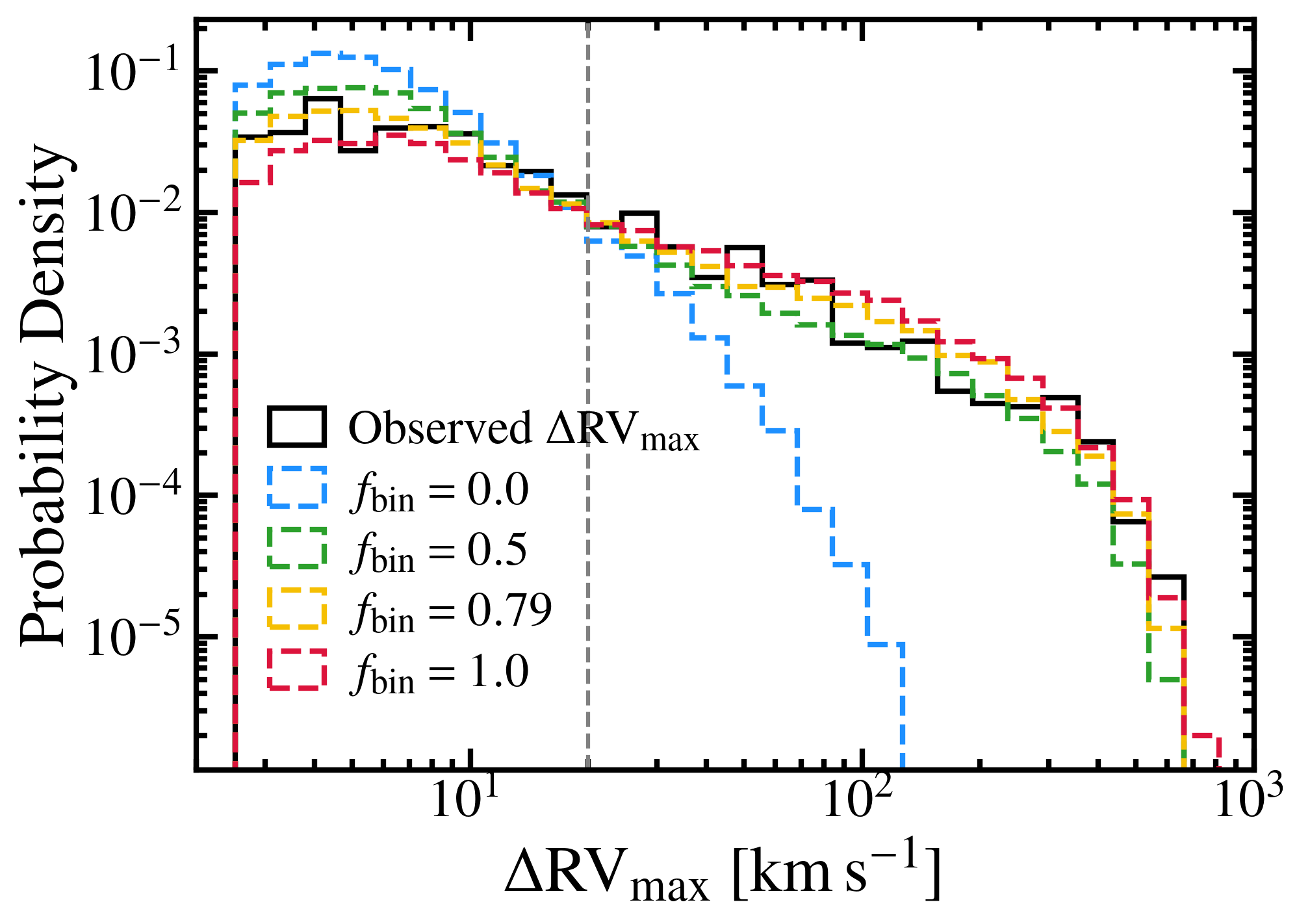}
    \caption{Distribution of \drvmax values for the BLOeM sample (black) and example simulated distributions for different binary fraction values, including our result of $f_{\rm bin}=0.79$. The dashed grey line indicates the 20\kms threshold.}
    \label{fig:dRVmax_distrib}
\end{figure}

The posterior distribution for $f_{\rm bin}$ is shown in Fig~\ref{fig:posterior_fbin}, our simulations clearly favour a high binary fraction of \fbin{79}{5} in good agreement with our value of \fbin{80}{8} reported in Sect.~\ref{sec:intrinsic_multp}. Figure~\ref{fig:dRVmax_distrib} shows the observed \drvmax distribution (black histogram), it can be seen that there is no evident separation between a core and tail component, possibly due to the uncertainty in our RV measurements or the scarcity of very wide binaries ($\log (P/{\rm d}) >4$) in comparison to the distribution of orbital periods of sun-like stars which peaks roughly at $10^5$\,d \citep[see][]{raghavan+10, badenes+18}. We include the distribution of \drvmax values for four simulated samples of 10\,000 stars computed with an intrinsic binary fraction of 0.0, 0.5, 0.79, and 1.0 to illustrate the change in the distribution with $f_{\rm bin}$. A lower binary fraction naturally increases the number of stars with low-amplitude RV variations, whereas a higher binary fraction adds more highly RV-variable systems. Notably, the transition value between single- and binary-dominated bins in the simulations occurs near 20\kms, supporting the choice of this value as a sensible threshold.

Another point worth mentioning is the inclusion of slightly evolved stars in our sample. As shown in Sect.~\ref{sec:mult_fract} and Fig.~\ref{fig:Pcrit}, for typical $q=0.5$ and $i=60^\circ$ detached systems with evolved primaries ($\log(g)=3.5$) can exhibit peak-to-peak RVs about 100\kms lower than those with $\log(g)=4.0$. These findings illustrate how stellar evolution can effectively shift some of the high-amplitude binaries towards lower values in the observed distribution of \drvmax. Because our simulations do not track orbital evolution with mass transfer or stellar ageing, these post-interaction systems might escape detection. Future \bloem papers will explore these evolutionary effects in detail. In conclusion, our multiplicity fraction can thus be interpreted as a lower limit due to binary evolution likely removing part of the high-\drvmax tail from the distribution.

\begin{figure}
    \centering
    \includegraphics[width=\linewidth]{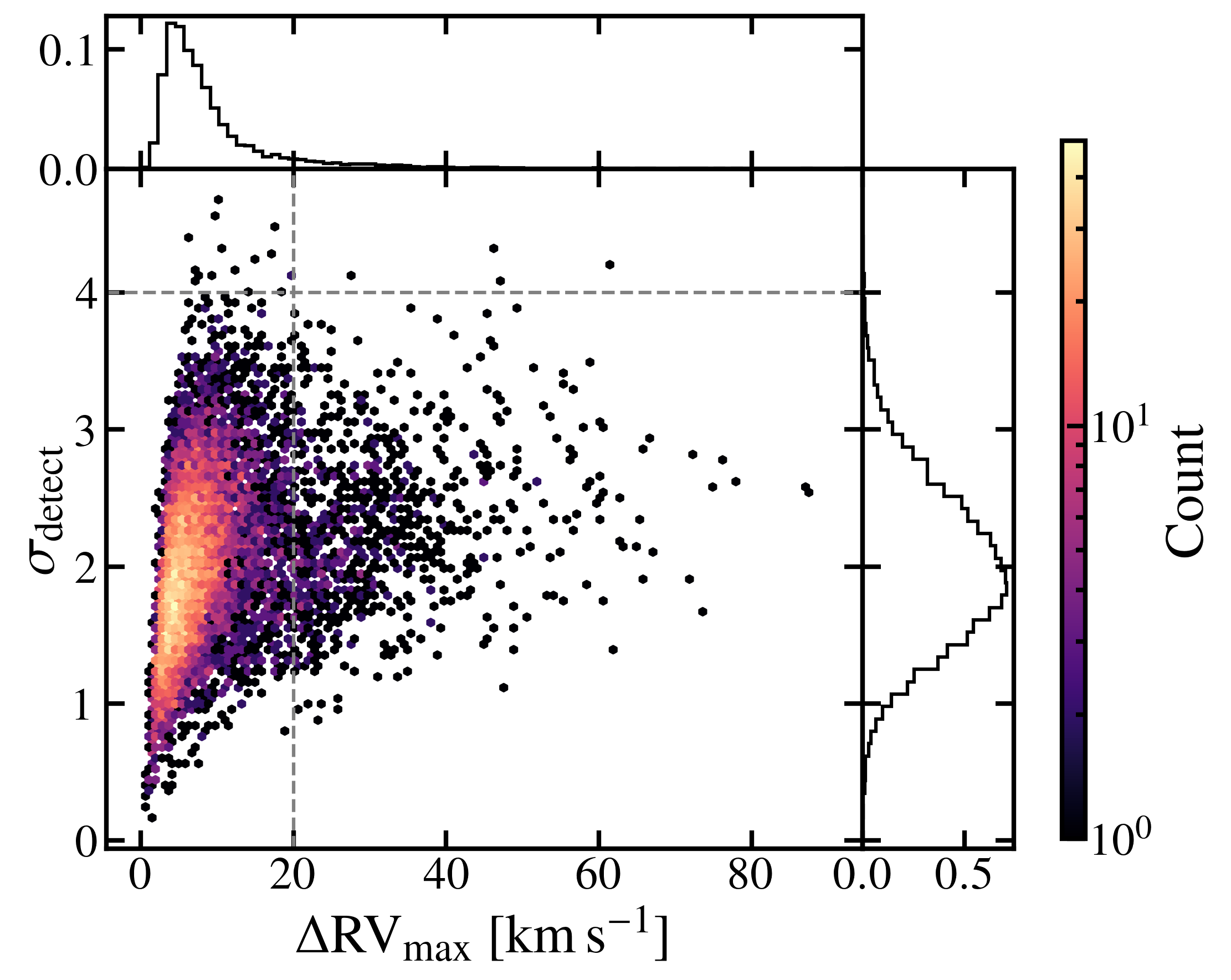}
    \caption{As Fig.~\ref{fig:sigma_d} for the simulated sample of single stars ($f_{\rm bin}=0$), colour coded by number of stars. Top and right panels show the 1D distribution of each of the binary criteria.}
    \label{fig:sigmad_dRV_sim}
\end{figure}

It is also interesting to examine the distribution of \drvmax values for $f_{\rm bin}=0$ (blue histogram in Fig.~\ref{fig:dRVmax_distrib}); there is a fraction of simulated single stars (9\%) with values  above 20\kms. The RV values are simulated by adding a Gaussian noise with a standard deviation corresponding to the uncertainty of the RV value at each epoch of a real, randomly assigned, \bloem star. One single bad epoch (e.g. large uncertainty due to low S/N or a bad multi-component fit) can produce a large \drvmax value. Figure~\ref{fig:sigmad_dRV_sim} shows the distribution of the simulated sample of single stars in the binary criteria diagram. Out of the 10\,000 stars, only four pass both binary criteria, meaning only 0.04\% false-positive detections. This exercise not only confirms the high multiplicity fraction of early B-type dwarfs and giants in the SMC, but also further validates the choice of the binary criteria for our sample.

\subsection{Variations in the orbital period distribution}\label{sec:porb_simulations}

\begin{figure}
    \centering
    \includegraphics[width=\linewidth]{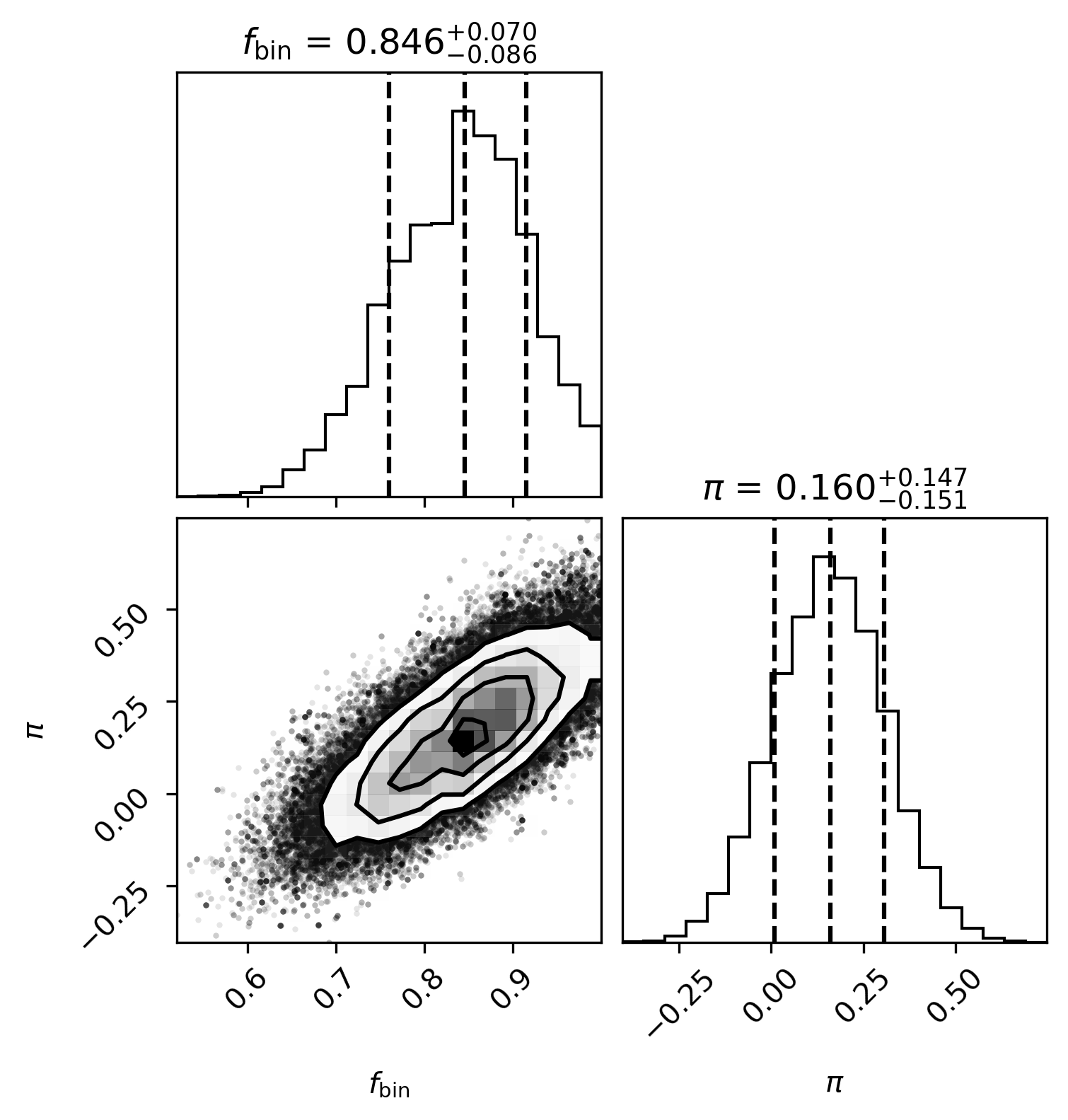}
    \caption{Corner plot illustrating the posterior distribution for the binary fraction ($f_{\rm bin}$) and period distribution exponent ($\pi$). The central (off-diagonal) panel displays the 2D joint posterior density, with contours indicating (2D) $1\sigma$, $2\sigma$, and $3\sigma$ confidence levels. The two marginal panels (diagonal) show the 1D distributions for $f_{\rm bin}$ and $\pi$, with vertical lines marking the 16\%, 50\%, and 84\% quantiles.}
    \label{fig:cornerplot}
\end{figure}

As mentioned in Sect.~\ref{ssec:RVthresh}, the distribution of orbital periods plays a significant role in the determination of the binary fraction. As a final test, we treat the exponent of the orbital period distribution $\pi$ as a free variable in our MCMC simulation. The resulting posterior distribution can be seen in Fig~\ref{fig:cornerplot}. Fitting both $\pi$ and $f_{\rm bin}$ simultaneously yielded $f_{\rm bin}=0.85^{+0.07}_{-0.09}$ and $\pi=0.16\pm0.15$. Although this represents a slight preference for longer periods ($\pi>0$) the result is essentially consistent with $\pi=0$. We find a significant correlation between both parameters, but overall, varying $\pi$ does not drastically alter the binary fraction, suggesting that while the precise shape of the period distribution is not tightly constrained by the current data, the inferred binary fraction remains robustly high. Once the full 25 epochs of observations from the survey become available, it will be possible to put much stronger constraints on the distribution of orbital periods.

\section{Conclusions} \label{sec:conclusions}

In this work, we focused on the B-type sample of dwarf and giant stars from the \bloem campaign in the SMC. We analysed 309 stars by measuring their RV variations to identify close binaries and constrain the multiplicity fraction. Our main conclusions are:

\begin{enumerate}
    \item High multiplicity fraction at low metallicity: We detect 153 spectroscopic binaries (91 SB1, 59 SB2, 3 SB3) based on significant RV variability. For this, we consider as binaries those systems which simultaneously fulfil having peak-to-peak RV variations above a threshold of $C=20$\kms and an RV significance above $4\sigma$, consistent with previous studies of massive stars in the MW and LMC. This corresponds to an observed multiplicity fraction of \fobs{50}{3}, exceeding reported values for B-type stars in the MW and LMC, and suggesting a potentially larger multiplicity fraction at lower metallicities.
    \item Orbital periods of the close binaries: Our multi-epoch strategy is well suited for identifying close systems with periods comparable to the total baseline of the observations. We obtained reliable orbital periods for close to 30\% of the detected binaries. However, we cannot confirm any significant differences with short-period binaries at higher metallicities. The full set of 25 \bloem observations will enable robust orbital solutions, providing a comprehensive understanding of the distribution of orbital parameters of our sample.
    \item Intrinsic multiplicity fraction from independent methods: After correcting for observational biases with MC simulations, we infer an intrinsic close-binary fraction of \fbin{80}{8}. Experimenting with a higher RV threshold reduces the multiplicity fraction down to $\sim55$\% for $C=80$\kms. However, an independent MCMC analysis, which does not rely on binary criteria but considers the full \drvmax distribution, returns a binary fraction of \fbin{79}{5}, confirming the high multiplicity. Letting the distribution of orbital periods vary further reaffirms our result, leading to \fbin{85}{7}{9}, although the orbital period distribution cannot be well constrained with our data.
    \item Potential anti-correlation with metallicity: The resulting multiplicity fraction is significantly higher in comparison to similar samples in the LMC and MW. This is consistent with an anti-correlation between metallicity and close-binary fraction among B-type III/V stars. While the fraction of O-type binaries seems to remain constant across different metallicities, B-type stars might be affected by how early or late disc fragmentation occurs during massive star formation. B-type binaries forming during extremely early fragmentation would have enough time and disc material to spiral inwards to very short orbits ($P_{\rm orb}<10$\,d). At later times however, low-metallicity discs have had more time to cool down due to the reduced opacity, increasing fragmentation and therefore enhancing multiplicity. However, with less accreting material, the star is not able to migrate to such low separations. This implies that the increased multiplicity fraction at low metallicity predominantly involves binaries with orbital periods above $\sim$10\,d, while the fraction of shorter-period binaries remains invariant with metallicity, consistent with previous studies of EBs.
    \item Implications for the early Universe: The SMC's metallicity is similar to that of many high-redshift, star-forming galaxies. A higher fraction of close binaries at low metallicity would translate to more frequent interactions, potentially increasing the number of exotic transients and envelope-stripped stars, relevant for cosmic reionisation. Our results also provide important input for population synthesis studies at low metallicity, and are critical for the population of gravitational-wave progenitors.
\end{enumerate}

\section*{Data availability}

Tables D.1 and D.2 are only available in electronic form at the CDS via anonymous ftp to cdsarc.u-strasbg.fr (130.79.128.5) or via \url{http://cdsweb.u-strasbg.fr/cgi-bin/qcat?J/A+A/}.

\begin{acknowledgements}
We thank the referee for his thoughtful and constructive comments that have greatly clarified our discussion and strengthened our conclusions. We acknowledge the support of the Data Science Group at the Max Planck Institute for Astronomy (MPIA) for their useful suggestions in the improvement of \texttt{ravel}.

The authors acknowledge support from the European Research Council for the ERC Advanced Grant 101054731.

DMB gratefully acknowledges funding from UK Research and Innovation (UKRI) in the form of a Frontier Research grant under the UK government's ERC Horizon Europe funding guarantee (SYMPHONY; grant number: EP/Y031059/1), and a Royal Society University Research Fellowship (URF; grant number: URF{\textbackslash}R1{\textbackslash}231631). 

IM acknowledges support from the Australian Research Council (ARC) Centre of Excellence for Gravitational Wave Discovery (OzGrav; project number CE230100016).

KS is funded by the National Science Center (NCN), Poland, under grant number OPUS 2021/41/B/ST9/00757. 

TS acknowledges support by the Israel Science Foundation (ISF) under grant number 0603225041

LRP acknowledges support by grants
PID2019-105552RB-C41 and PID2022-137779OB-C41 funded by
MCIN/AEI/10.13039/501100011033 by "ERDF A way of making
Europe". 
LRP acknowledges support from grant PID2022-140483NB-C22 funded by MCIN/AEI/10.13039/501100011033.

DP acknowledges financial support from the FWO junior postdoctoral fellowship No. 1256225N. 

MP is supported by the BEKKER fellowship BPN/BEK/2022/1/00106 from the Polish National Agency for Academic Exchange and the Royal Physiographic Society in Lund through the Märta and Erik Holmbergs Endowment.
\end{acknowledgements}

\bibliographystyle{aa} 
\bibliography{JVbiblio.bib}

\begin{appendix}

\section{Additional figures}\label{append:figs}
\subsection{Spatial distribution of the B III/V sample}\label{append:spatial}

\begin{strip}
   \centering
   \includegraphics[width=\hsize]{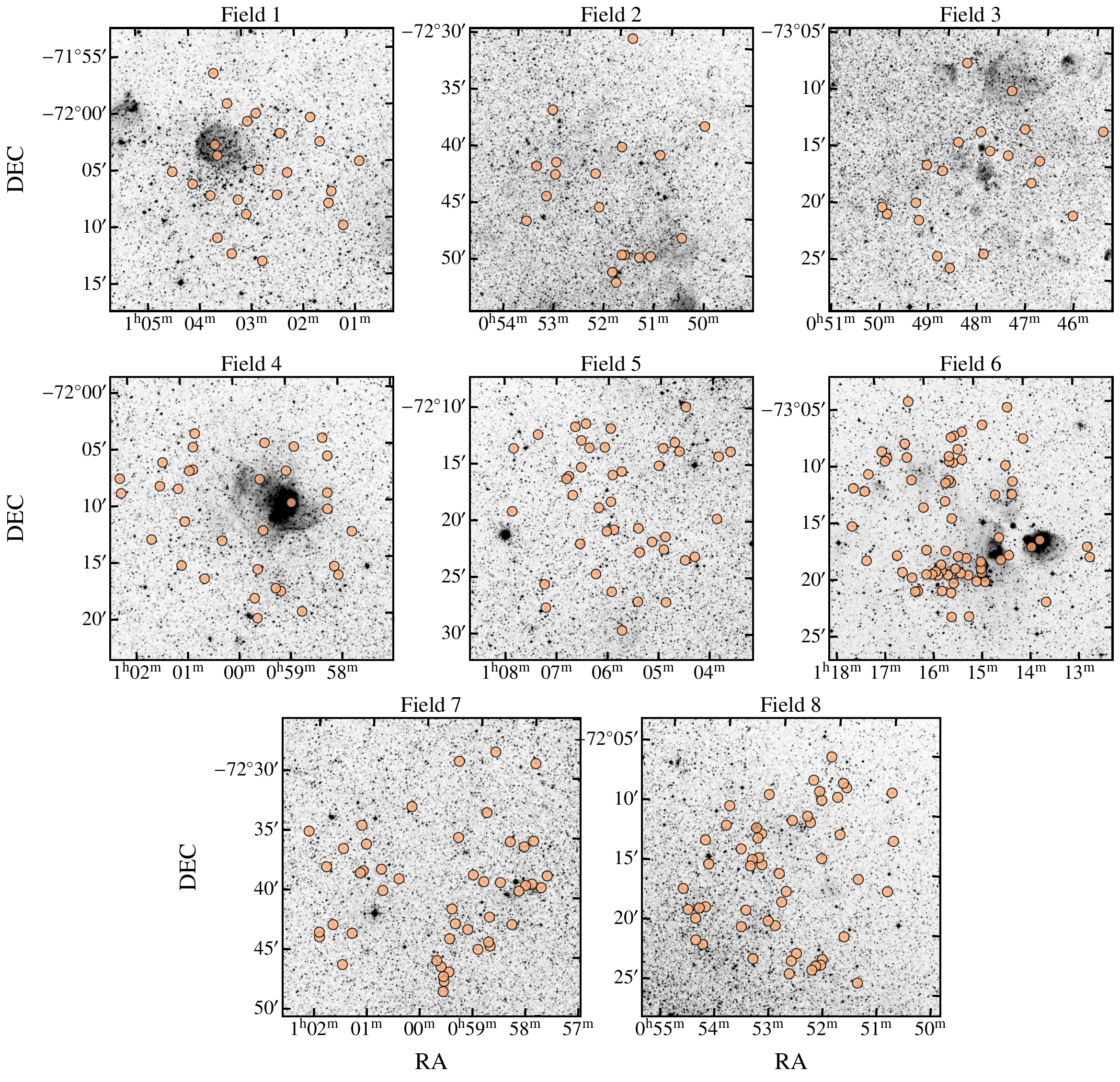}
      \captionof{figure}{Spatial distribution of the B-type dwarfs across the eight fields observed by the \bloem campaign. The bright star-forming region NGC~346 is featured in Field 4, however, it has not been extensively covered by the \bloem campaign (see main text for details).}
      \label{fig:smc_fields}
\end{strip}

\subsection{Hertzsprung-Russell diagram}\label{append:HRD}

\begin{figure*}
   \centering
   \includegraphics[width=\hsize]{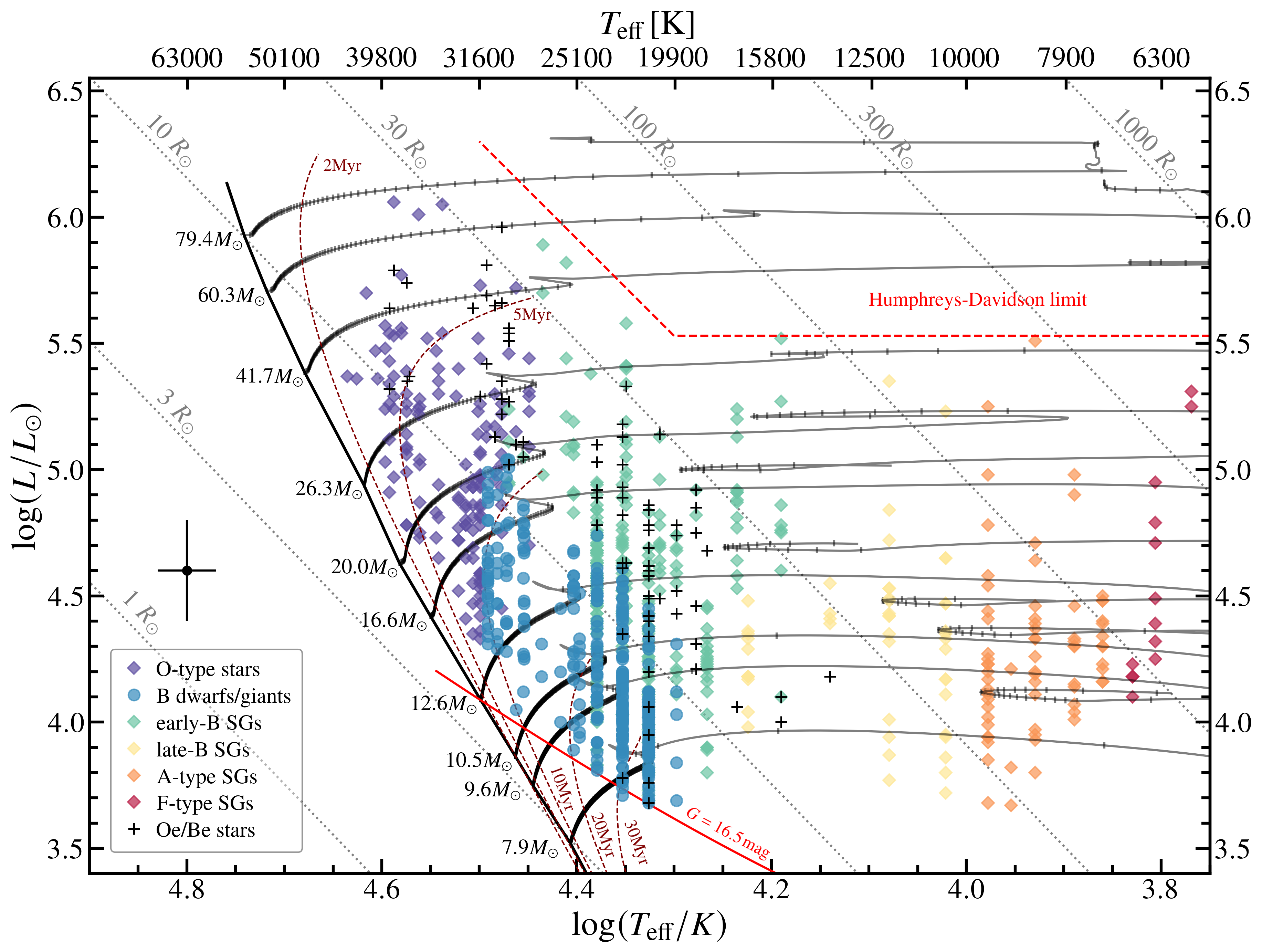}
      \caption{Hertzsprung–Russell diagram of the full \bloem sample. Stars are colour-coded by spectral type, as indicated in the legend, and \Teff and $\log(L)$ values are from Bestenlehner et al. (in prep.). The black curves show evolutionary tracks for different initial masses at SMC metallicity from \citet{schootemeijer+19} with an overshooting prescription from \citet{hastings+21}. Lines of constant radius are marked in grey, and isochrones of different ages are shown with dashed brown curves. The black cross on the left illustrates typical uncertainties in \Teff and $\log(L)$.}
      \label{fig:hrd_full}
\end{figure*}

Figure \ref{fig:hrd_full} shows the Hertzsprung-Russell diagram (HRD) of the full \bloem sample, colour-coded by their different spectral type regimes. Each of the subsamples has been investigated independently: the O-type stars by \citet{sana+24}, the B-type dwarfs and giants in this work, the early B-type supergiants (BSGs) by \citet{britavskiy+25}, the Oe/Be stars by \citet{bodensteiner+25}, and the late BAF supergiants by \citet{patrick+25}. All these works have investigated the multiplicity fraction of their respective samples, and there is a clear trend with evolutionary phase. The two samples with the less evolved content, non-supergiant O- and B-type stars, have the highest multiplicity fractions, with $70^{+11}_{-6}\%$ found by \citet{sana+24} and $80\pm8\%$ found in this work. The fraction rapidly decreases for early BSGs, reaching $40\pm4\%$. Furthermore, \citet{britavskiy+25} reports a drop in the observed binary fraction at spectral types later than B2. Looking at the HRD of Fig.~\ref{fig:hrd_full}, we can see that the cooler BSGs have $\log (L/L_\odot)$ values above 4.5\,dex. Therefore, it is likely that the drop in the binary fraction is related to the evolutionary phase of these stars; namely, the large radii of the BSGs have induced binary interactions, effectively removing the close binaries. This is supported by the study of \citet{bodensteiner+25}, who found a binary fraction of 18-32\% among the emission-line stars. The stark contrast in the multiplicity properties of Oe/Be stars and OB-type stars strongly suggests that the former are mostly binary interaction products, as proposed by \citet{bodensteiner+25}. In many cases, the initially more massive star has lost a large fraction of its envelope mass and produces RV variations on the now more massive companion that are too small to detect, leading to a lower observed binary fraction. The more evolved stars in the \bloem sample, the BAF supergiants, present the lowest binary fraction of all, with less than $18\%$ of the late BSGs and $8^{+9}_{-7}\%$ of the A- and F-type stars showing binary signatures \citep{patrick+25}. Although the intrinsic binary fractions for the supergiant samples are highly uncertain due to evolutionary aspects and the unclear role of intrinsic variability, the trend is clear. Binary evolution strongly impacts the evolutions of massive stars, as is evidenced by the large drop in the binary fraction from the MS to the late supergiant phases; if the intrinsic binary fraction of AF supergiants is really below 20\%, then close to 60\% of massive stars must have gone through envelope stripping or mergers before reaching the AF supergiant phase, and now appear as single stars.

\section{Radial velocity uncertainties}\label{append:rverr}

Due to the importance of the RV uncertainties for the identification of binaries and ultimately for the intrinsic binary fraction, we have investigated the reliability of the RV uncertainties obtained form our spectral line profile fitting approach. In order to compare with a different and independent technique, we have performed cross-correlation in a similar way to that implemented by \citet{mahy+22} and also applied to other \bloem samples (\citet{bodensteiner+25}, \citet{patrick+25}, \citet{britavskiy+25}). Due to the difficulties of cross correlation handling SB2 systems, we exclude the SB2 and SB3 systems from the comparison.

   \begin{figure}
   \centering
   \includegraphics[width=\hsize]{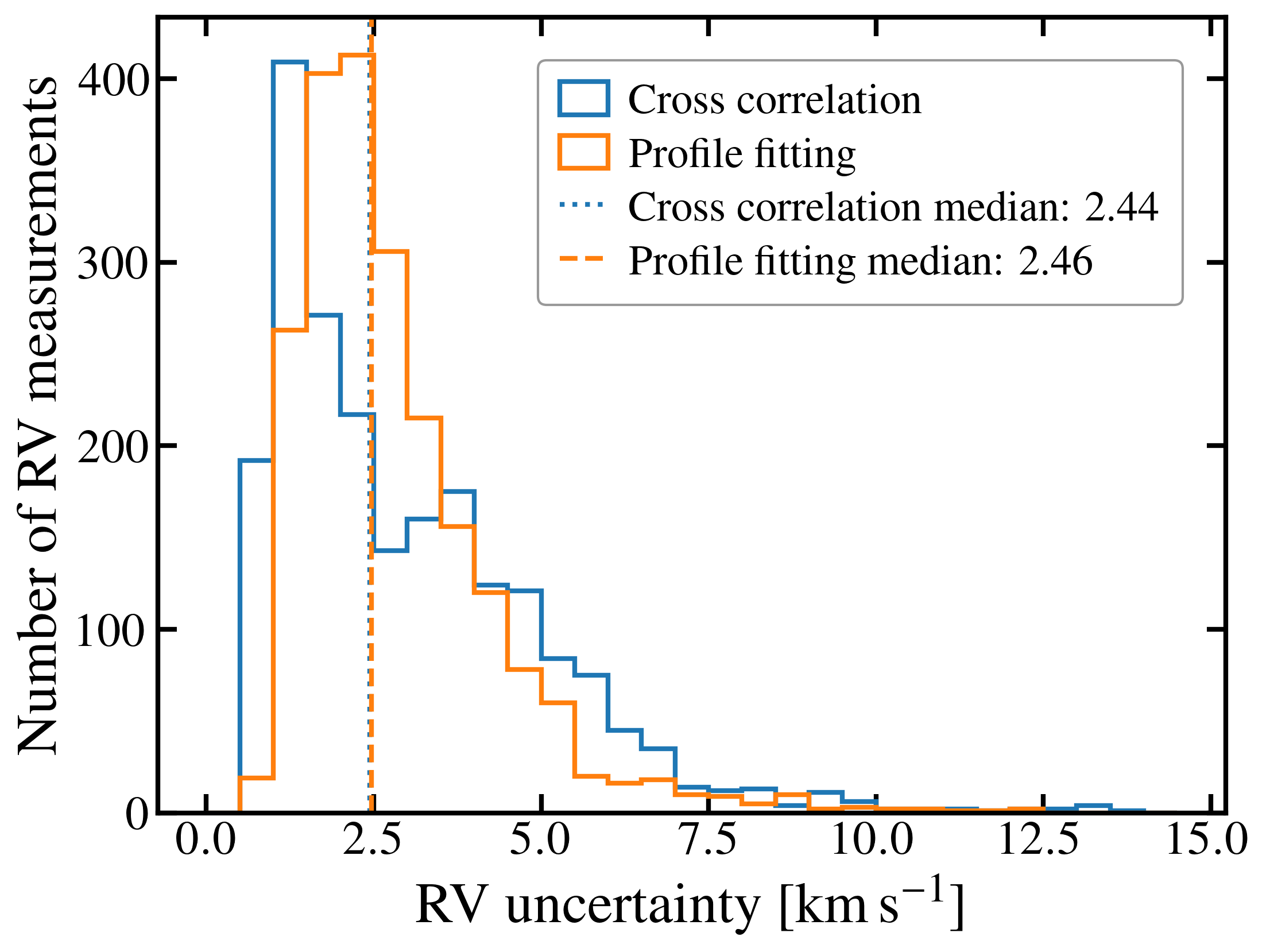}
      \caption{Comparison of the uncertainties on the individual RV measurements provided by our profile fitting technique and those from cross correlation.}
      \label{fig:RVerr}
   \end{figure}

The individual RV comparison is shown in the histogram of Fig.~\ref{fig:RVerr}. In both cases there is a tail after 7\kms which goes to almost zero after 10\kms. The cross correlation uncertainties have a primary peak between 1-1.5\kms, and a smaller secondary peak close to 4\kms,  whereas the uncertainties of the profile fitting technique peak at 2\kms. However, both distributions have nearly identical medians close to 2.45\kms. This simple comparison shows that there is no underestimation of our RV uncertainties that could lead to an overestimation of $\sigma_{\rm d}$, and consequently to a larger multiplicity fraction.

   \begin{figure}
   \centering
   \includegraphics[width=\hsize]{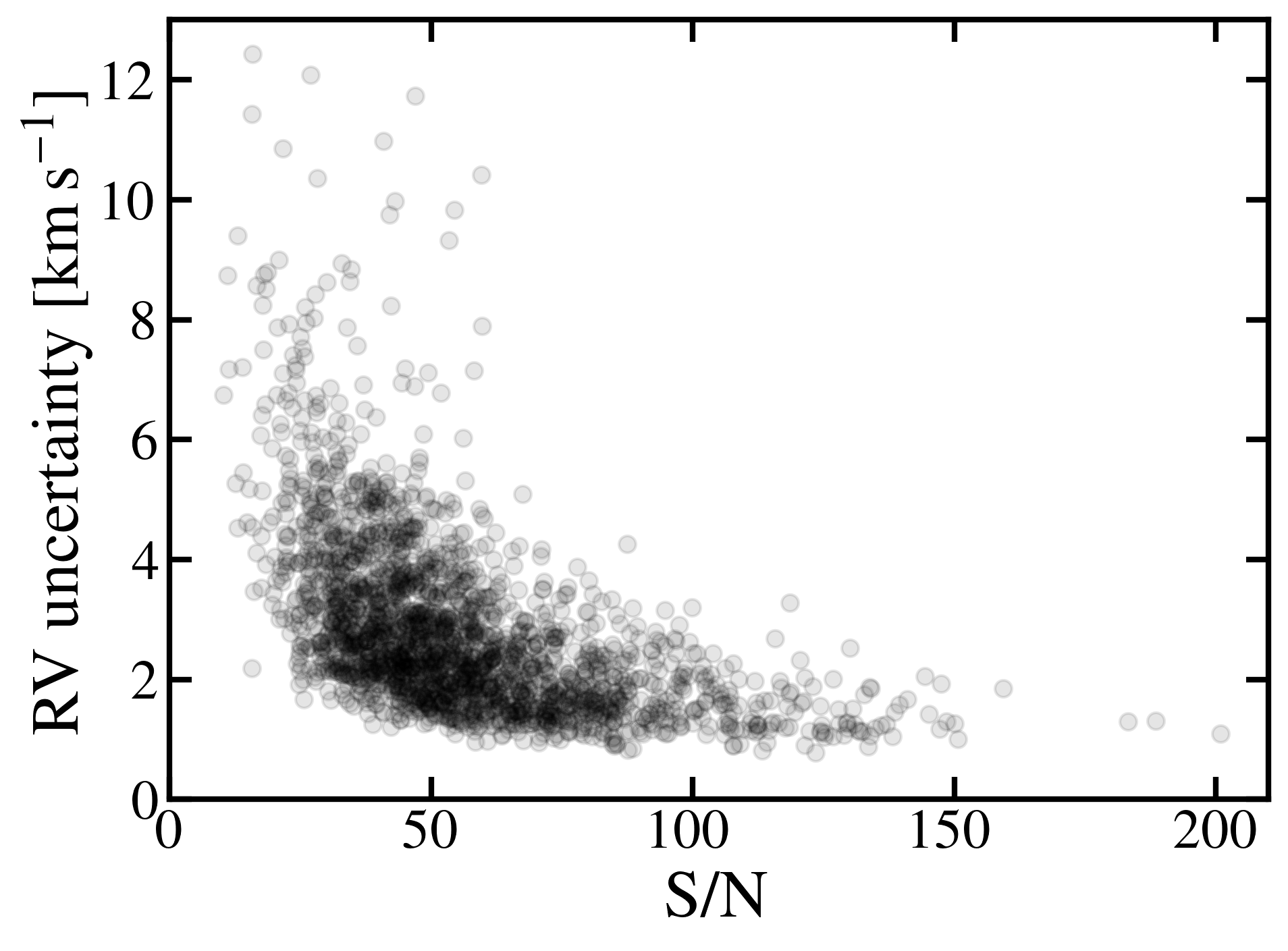}
      \caption{Radial velocity uncertainty of our profile fitting approach as a function of S/N of the individual epochs.}
      \label{fig:RVerr_s2n}
   \end{figure}

In Fig.~\ref{fig:RVerr_s2n} we show the dependence of our RV uncertainty on the S/N of the individual observations. We can clearly see how the uncertainty anti-correlates with S/N. Spectra with S/N > 70 consistently have RV uncertainties of less than 3\kms. Spectra with S/N < 50 have a larger spread on uncertainties, but most of them have uncertainties below 5\kms.

\section{Notes on individual systems}\label{append:indiv_sys}

\subsection{Confirmed SB3 systems}\label{append:sb3}

\begin{figure}
   \centering
   \includegraphics[width=\hsize]{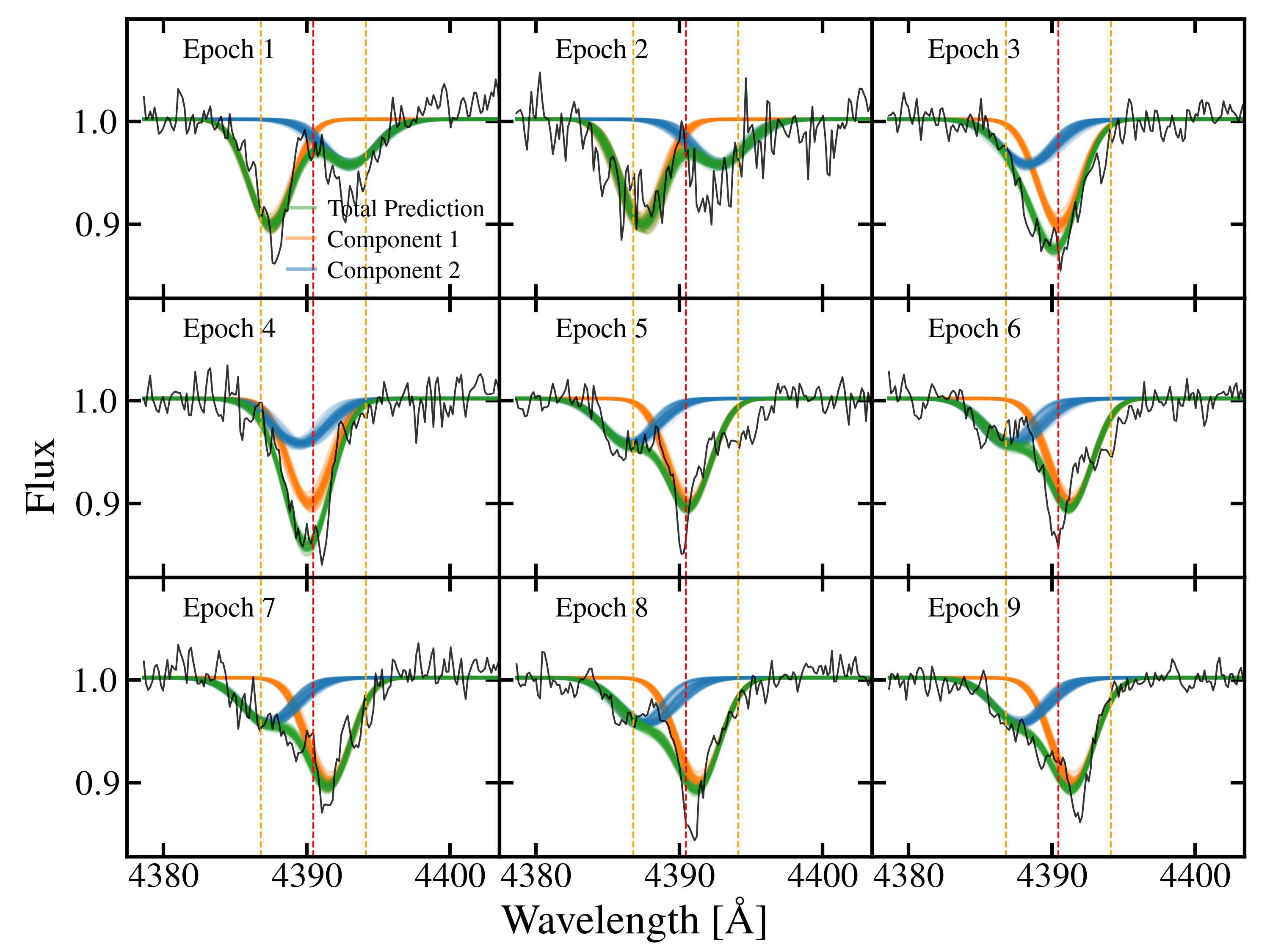}
      \caption{Gaussian fitting of the spectral lines \spline{He}{i}{4388} for the SB3 system \bloem 5-062. The dashed red line shows the mean SMC velocity of 172\kms \citep{evans+howarth08}, and the dashed orange lines show differences of $\pm200$\kms with respect to the mean velocity. Epoch number increases from left to right and from top to bottom.}
      \label{fig:5_062_SB3}
   \end{figure}

\paragraph{\bloem 5-062:} (OGLE-SMC-ECL-5096) Characterised by a narrow absorption feature moving with a large RV. As shown in Fig.~\ref{fig:5_062_SB3}, the core of the line is not well fitted with a Gaussian profile, and in epochs 5 and 6 there are two other absorption features at both sides of the narrow absorption. Our inspection of the photometric data shows that this is likely a doubly eclipsing system, with a potential outer period of 77.2\,d, close to the spectroscopic period (80\,d). The OGLE (inner) photometric period is 5.2\,d.

   \begin{figure}
      \centering
      \includegraphics[width=\hsize]{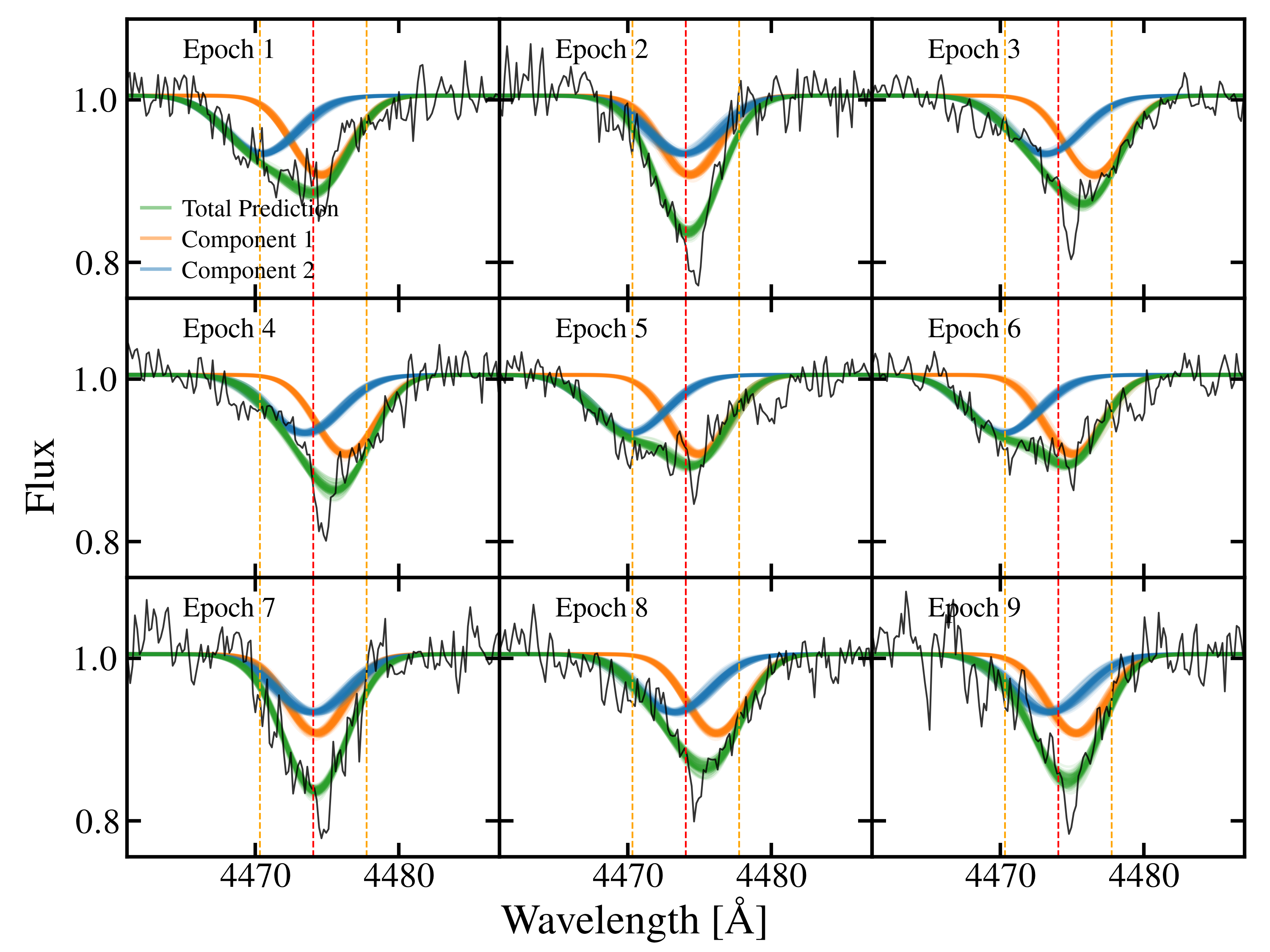}
         \caption{As Fig.~\ref{fig:5_062_SB3}, this time displaying line \spline{He}{i}{4471} for \bloem 6-062.}
         \label{fig:6_062_SB3}
   \end{figure}

\paragraph{\bloem 6-062:} (OGLE-SMC-ECL-5853) Similarly to \bloem 5-062, it shows an absorption feature with a narrow core, but in this case the core seems stationary or moving at a much lower velocity as it can be seen in Fig.~\ref{fig:6_062_SB3}, suggesting it could belong to the component in the wider orbit. The amplitude of the narrow feature varies drastically among epochs, a sign of the movement of an inner pair. The two other absorption features are more clearly visible in epoch 4. The OGLE period is 1.13\,d, whereas the spectroscopic period is 1.98\,d, which we did not consider significant, and it is clearly affected by the contribution of the third component.

\begin{figure}
   \centering
   \includegraphics[width=\hsize]{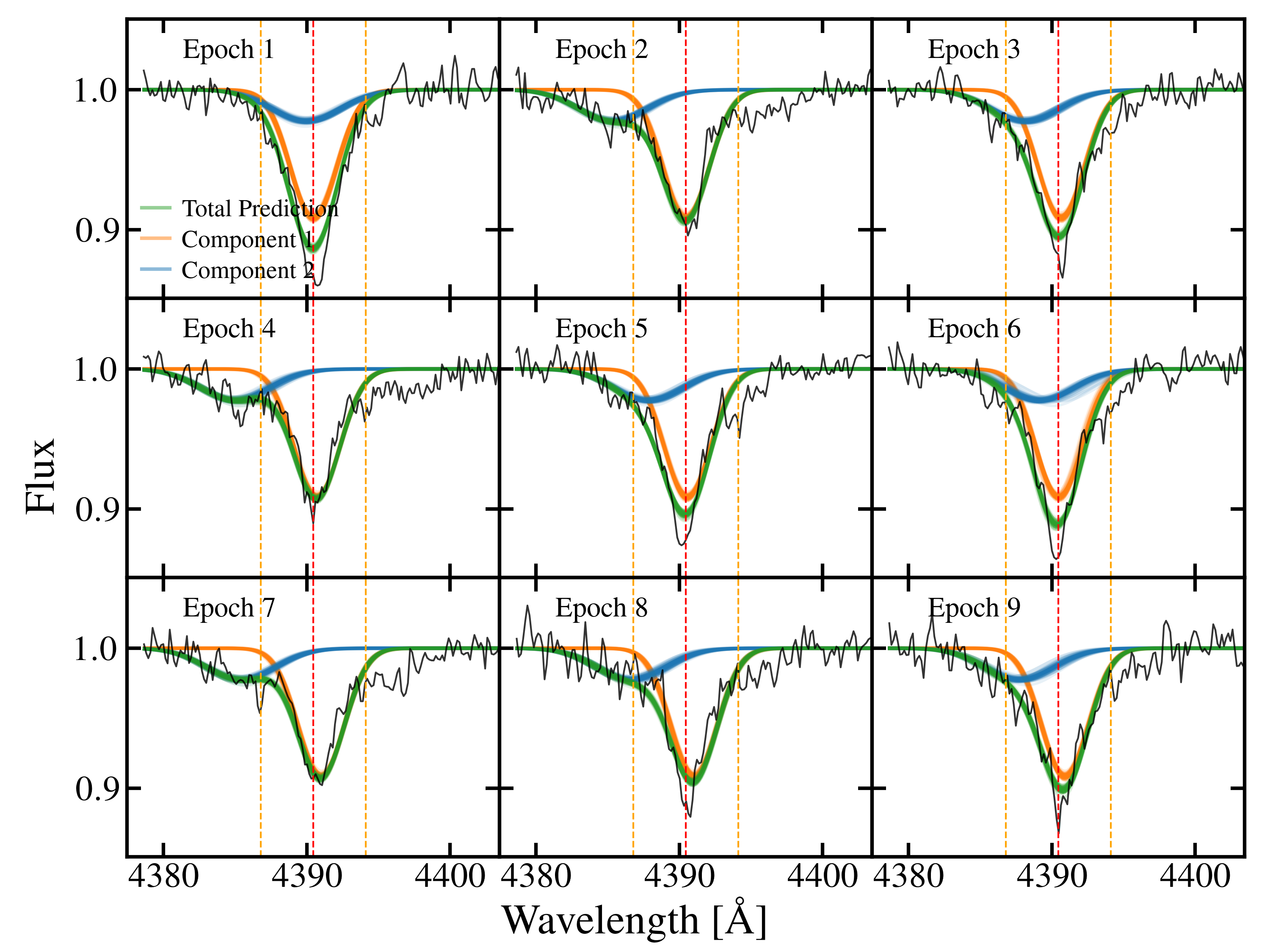}
      \caption{As Fig.~\ref{fig:5_062_SB3} for \bloem 7-057.}
      \label{fig:7_057_SB3}
\end{figure}

\paragraph{\bloem 7-057:} (OGLE-SMC-ECL-4024) As shown in Fig.~\ref{fig:7_057_SB3}, it clearly displays a narrower component, apparently stationary, and a broader one that is difficult to identify as two components. However, the profile variability in the broader component is evident, and consistent with large RV amplitudes of an inner pair, specially in epochs 2, 4, and 7. The photometric data suggest this is possibly a doubly eclipsing system, and a possible secondary period of 2.57\,d (half the low-significance spectroscopic period).

\subsection{Other eclipsing binaries}\label{append:EBs}

\paragraph{\bloem 1-037:} (OGLE-SMC-ECL-4510) Wrong period in the OGLE catalogue (1/3 harmonic), actual period of 9.975\,d (twice the spectroscopic period).

\paragraph{\bloem 2-051:} (OGLE-SMC-ECL-2251) Eccentric orbit with apsidal motion, possibly due to a third companion. However, no evidence of a double eclipse was found.

\paragraph{\bloem 3-050:} (OGLE-SMC-ECL-6645) Possibly doubly eclipsing. No signs of outer period though.

\paragraph{\bloem 3-072:} (OGLE-SMC-ECL-1475) Possibly a doubly eclipsing system, possible secondary period of 10.03\,d. This system was flagged as possible SB3.

\paragraph{\bloem 4-095:} (OGLE-SMC-ECL-4255) Probably doubly eclipsing. Secondary OGLE period of 2.72\,d, also identified spectroscopically. This system was flagged as possible SB3.

\paragraph{\bloem 5-002:} (OGLE-SMC-ECL-4756) Probably doubly eclipsing. Secondary OGLE period of 2.06\,d (spectroscopic period).

\paragraph{\bloem 5-030:} (OGLE-SMC-ECL-7623) Spectroscopic period is half of the photometric period.

\paragraph{\bloem 6-115:} (OGLE-SMC-ECL-5933) Spectroscopic period is twice the photometric period.

\paragraph{\bloem 7-032:} (OGLE-SMC-ECL-3848) Probably doubly eclipsing. Secondary OGLE period of 17.58\,d (slightly longer than the spectroscopic period of 16.66\,d). This system was flagged as possible SB3.

\paragraph{\bloem 7-060:} (OGLE-SMC-ECL-4028) Spectroscopic period is half of the photometric period.

\paragraph{\bloem 8-029:} (OGLE-SMC-ECL-2377) Spectroscopic period is twice the photometric period.

\paragraph{\bloem 8-105:} (OGLE-SMC-ECL-2967) Spectroscopic period is a fourth of the photometric period. Eccentric orbit.

\onecolumn
\section{Tables}\label{append:tables}
\FloatBarrier  

\input{tables/binaries}

\input{tables/eclipsing_binaries}

\end{appendix}
\end{document}

%% file: tables/binaries.tex
\footnotesize
\begin{longtable}{llrrrlrllll}
  \captionsetup{width=\linewidth} \\
  \caption{Multiplicity properties of the \bloem sample of B-type dwarfs and giants. Column “n'' corresponds to the number of epochs available or used in RV determination. Column “P\_sig'' indicates whether the LS orbital period is significant (S), tentative (T) or not significant (NS), as defined in Sect.~\ref{sec:bloem_periods}. Column “EB'' highlights eclipsing binaries from OGLE, and “OGLE\_P'' is the OGLE orbital period.}\\
  \label{tab:bin} \\
  \toprule
  \toprule
  BLOeM~ID &            SpT &  n &     $\Delta$RV &  $\sigma_{\rm d}$ &   Binarity &  Period & P\_sig & EB &     OGLE\_P &                             Comments \\
  & & & \kms & & & [d] & & & [d] & \\
  \midrule
  \endfirsthead
  \caption{continued.}\\
  \toprule
  \toprule
  BLOeM~ID &            SpT &  n &     $\Delta$RV &  $\sigma_{\rm d}$ &   Binarity &  Period & P\_sig & EB &     OGLE\_P &                             Comments \\
  & & & \kms & & & [d] & & & [d] & \\
  \midrule
  \endhead
  \midrule
  \endfoot
  BLOeM\_1-002 &         B0 IV: &        9 &   66.33 &    31.21 &   SB1 &   48.87 &     T & -- &         -- &                                  -- \\
  BLOeM\_1-010 &      B1.5 III: &        9 &    9.76 &     3.17 & RVvar &    1.48 &    NS & -- &         -- &                                  -- \\
  BLOeM\_1-013 &      B1.5 III: &        9 &   69.94 &    12.29 &   SB1 &    2.61 &     T & -- &         -- &                                  -- \\
  BLOeM\_1-015 &       B0.5 III &        9 &   74.14 &    31.09 &   SB1 &   22.11 &    NS & -- &         -- &                                  -- \\
  BLOeM\_1-020 &         B0 III &        9 &   93.23 &    41.72 &   SB1 &   38.34 &     T & -- &         -- &                                  -- \\
  BLOeM\_1-022 &         B1 III &        9 &    8.52 &     1.70 & RVcst &    1.29 &    NS & -- &         -- &                                  -- \\
  BLOeM\_1-032 &      B1.5 III: &        9 &  121.15 &    38.40 &   SB1 &    3.05 &     S & -- &         -- &                                  -- \\
  BLOeM\_1-035 &    B1.5 III-II &        9 &    8.60 &     2.96 & RVvar &    2.04 &    NS & -- &         -- &                                  -- \\
\end{longtable}

%% file: tables/eclipsing_binaries.tex
\begin{table*}
\centering
\caption{Summary of Eclipsing Binaries}
\label{tab:eclipsing_binaries}
\begin{tabular}{l l l c c l}
\toprule
   BLOeM ID & OGLE ID & Spectral Type &  $I$-band &  Period & Type \\
    & & & [mag] &  [d] & \\
\midrule
1-037 & OGLE-SMC-ECL-4510 &      B1.5 III &           15.65 &        3.32 &   NC \\
2-051 & OGLE-SMC-ECL-2251 &      B2.5 IV: &           16.26 &        2.34 &   NC \\
2-052 & OGLE-SMC-ECL-2272 &       B0.5 V: &           14.05 &        3.82 &   NC \\
2-064 & OGLE-SMC-ECL-2419 &   B1.5: V-III &           15.06 &        2.58 &   NC \\
3-050 & OGLE-SMC-ECL-6645 &      B1.5 III &           15.04 &        4.42 &   NC \\
3-067 & OGLE-SMC-ECL-1466 &       B0.2 IV &           14.92 &        2.00 &   NC \\
\bottomrule
\end{tabular}
\tablefoot{Types are NC: detached or semi-detached, C: contact, ELL: ellipsoidal.}
\end{table*}